\documentclass[prd,aps,a4paper,twocolumn,eqsecnum,nofootinbib,floatfix]{revtex4}  %

\usepackage{graphicx} 
\usepackage{amsmath,amsfonts,amssymb}
\usepackage{color}
\newcommand{\beq}{\begin{equation}}
\newcommand{\eeq}{\end{equation}}
\newcommand{\bea}{\begin{eqnarray}}
\newcommand{\eea}{\end{eqnarray}}
\newcommand{\pinf}{p_\infty}

\begin{document}

\title{Radiation-reaction and angular momentum loss at the second Post-Minkowskian order}

\author{Donato Bini$^{1,2}$, Thibault Damour$^3$}
  \affiliation{
$^1$Istituto per le Applicazioni del Calcolo ``M. Picone,'' CNR, I-00185 Rome, Italy\\
$^2$INFN, Sezione di Roma Tre, I-00146 Rome, Italy\\
$^3$Institut des Hautes Etudes Scientifiques, 91440 Bures-sur-Yvette, France
}

\date{\today}

\begin{abstract}
We compute the variation of the Fokker-Wheeler-Feynman   total linear and angular momentum of a gravitationally interacting binary system  under the second post-Minkowskian retarded dynamics.
The resulting $O(G^2)$ equations-of-motion-based,  total change in the system's angular momentum is found to agree with existing computations that assumed  balance with angular momentum fluxes in the radiation zone.  
 \end{abstract}

\maketitle

\section{Introduction}

The issue of angular momentum loss during the scattering of two gravitationally interacting  particles has recently attracted a lot of attention~\cite{Damour:2020tta,DiVecchia:2021ndb,Herrmann:2021lqe,Jakobsen:2021smu,Mougiakakos:2021ckm,Compere:2021inq,DiVecchia:2021bdo,Herrmann:2021tct,Jakobsen:2021lvp,Bini:2021gat,Saketh:2021sri,Gralla:2021qaf,Veneziano:2022zwh,Manohar:2022dea,Alessio:2022kwv,DiVecchia:2022nna,Riva:2022fru,Kalin:2022hph,Chen:2022fbu,DiVecchia:2022piu,Heissenberg:2022tsn,Damour:2022ybd,Porrati2022}.
The existing post-Minkowskian(PM)-accurate computations of angular momentum loss have relied on an assumed balance between the angular momentum of the system of two point masses and the radiative fluxes of angular momentum at future null infinity. 
However, several authors have emphasized that the super-translation dependence of the definition of angular momentum at infinity raises concerns about using such balance laws for deriving effects related to the change of  mechanical state of the two particles during scattering~\cite{Veneziano:2022zwh,Porrati2022}. Another puzzling feature of having an
angular momentum loss at $O(G^2)$ \cite{Damour:2020tta}, while the energy-momentum loss starts at 
$O(G^3)$ \cite{Herrmann:2021lqe,Westpfahl:1987hwd}, is the resulting apparent
clash with the idea that, in a quantum computation, all radiative losses at infinity are a priori expected to involve
the emission of real gravitons \cite{Veneziano:2022zwh}.

The aim of the present work is to show that it is possible to by-pass any  ambiguity concerning radiative losses of angular momentum by using an approach entirely based on the retarded equations of motion of the binary system. The possibility to do so had been first demonstrated many years ago, at the lowest Post-Newtonian (PN) order, 2.5PN, in Ref. \cite{Damour:1981bh}.
In the latter reference the 2PM retarded equations of motion (in harmonic coordinates)  of a binary system derived in Ref. \cite{Bel:1981be} were PN-expanded up to the  $(v/c)^5$-accuracy. It was then shown that the 2PN ($(v/c)^4$) truncation of the equations of motion described a Poincar\'e-invariant conservative dynamics admitting ten Noetherian conserved quantities, namely total linear momentum and total angular momentum \cite{[23]81}.

The total variation, under the retarded dynamics, of the Noetherian linear momentum and  angular momentum during scattering were then computed. This led, in particular, to a variation during scattering of the Noetherian angular momentum of the system of order $G^2/c^5$.

In the present work we extend this logic to the PM framework, without ever making use of PN expansions. More precisely, 
we  compute the variation of the Fokker-Wheeler-Feynman \cite{Fokker:1929,Wheeler:1949hn}   total linear and angular momentum of the  binary system~\cite{Dettman:1954zz,Friedman:2005rx}  
under the exact (harmonic-coordinates) 2PM retarded dynamics \cite{Bel:1981be,Westpfahl:1979gu}. On the one hand, we find (confirming previous results \cite{Westpfahl:1987hwd}) that the variation of the total linear momentum vanishes at order $G^2$. On the other hand, our retarded-equations-of-motion-based computation  of the total change in the Noetherian angular momentum  of the system (using Lagrange's method of variation of constants) 
leads to a nonzero $O(G^2)$ result which agrees with existing computations (starting with Ref. \cite{Damour:2020tta}) that relied on computing fluxes of angular momentum at future null infinity.

We view our present  study as the first step in an approach which can in principle be extended to higher PM levels. This might bring additional light on the present puzzles that affect the understanding of the 5PN dynamics
 \cite{Blumlein:2021txe,Bini:2021gat,Almeida:2022jrv}. See \cite{Dlapa:2022lmu,Bini:2022enm}
  for a recent clarification of radiation-reaction effects at the $O(G^4)$ level.

\section{Retarded force at $O(G^2)$}

We consider two gravitationally interacting  point masses, $m_1$ and $m_2$, with world lines ${\mathcal L}_1$ and ${\mathcal L}_2$ having proper time\footnote{We use Minkowski proper time $d\tau_a=\sqrt{-\eta_{\alpha\beta}dz_a^\alpha dz_a^\beta}$ in a  mostly plus signature, $\eta_{\alpha\beta}={\rm diag}[-1,1,1,1]$.} parametric equations  $z_1^\mu(\tau_1)$  and $z_2^\mu(\tau_2)$.
The Poincar\'e-invariant  {\it retarded} equations of motion of the world lines have been explicitly derived (in harmonic coordinates) at the order $G^2$ (second Post-Minkowskian approximation, 2PM) in Refs. \cite{Westpfahl:1979gu,Bel:1981be}. They read ($a,b=1,2$, with $a\not = b$)
\beq
\label{master_eq}
m_a\frac{d^2z_a^\mu(\tau_a)}{d\tau_a^2} =F_{a\rm R}^\mu[z_a(\tau_a),u_a(\tau_a); z_{b\rm R}(\tau_a), u_{b\rm R}(\tau_a)] \,,
\eeq
where 
\beq
\label{wl_dep_force}
F_{a\rm R}^\mu[z_a,u_a; z_{b\rm R}, u_{b\rm R}]=F_{a\rm R\,, 1PM}^\mu+F_{a\rm R\,, 2PM}^\mu+O(G^3)\,.
\eeq
Here the label ``R"    stands for retarded,  $z_{b\rm R}^\mu(\tau_a)=z_b^\mu[\tau_{bR}(\tau_a)]$ is the retarded \lq\lq pre-image" of  $z_a^\mu(\tau_a)$ on ${\mathcal L}_b$,
\beq
(z_a(\tau_a)-z_{b}[\tau_{bR}(\tau_a)])^2=0\,\qquad z_a^0-z_{b\rm R}^0>0\,,
\eeq
and  $u_a^\mu(\tau_a)=\frac{dz_a^\mu(\tau_a)}{d\tau_a}\equiv \dot z_a$ ($a=1,2$),   $u_{b\rm R}^\mu(\tau_a)=u_b^\mu(\tau_b)\big|_{\tau_b=\tau_{bR}(\tau_a)}$.

The explicit expressions of the accelerative forces  $\Gamma_{a\rm R, 1PM}^\mu \equiv\frac{1}{m_a}F_{a\rm R, 1PM}^\mu$ and  $\Gamma_{a\rm R, 2PM}^\mu \equiv \frac{1}{m_a}F_{a\rm R, 2PM}^\mu$ are, for particle $m_1$ (see Eqs. (118)-(133) in Ref. \cite{Bel:1981be})
\bea
\label{gamma_1_2}
\Gamma_{1   \rm R\,, 1PM}^\alpha&=&\frac{Gm_2}{\rho_R^2}[(1-2\omega_R^2)A_R^\alpha\nonumber\\
& -&(1+2\omega_R^2 +4A_R\omega_R)v_R^\alpha] \,,\nonumber\\
\Gamma_{1 \rm R\,, 2PM}^\alpha 
&=& \frac{G^2}{\rho_R^3}[m_2^2 (a_0 A_R^\alpha +c_0 v_R^\alpha) +m_1m_2 (aA_R^\alpha +c v_R^\alpha)]\nonumber\\
&-&4m_1 \frac{d}{ds} [\Gamma_{1\, R}^\alpha \ln A_R]\,,
\eea
with
\bea
\label{ac_coeffs}
a_0&=& 2  [2\omega_R^2+(\omega_R+A_R)^2]\,,\nonumber\\
c_0&=&-2 [2\omega_R^2 +A_R (\omega_R+A_R)]\,, \nonumber\\
a&=& -\frac{2}{A_R^5}-\frac{5\omega_R}{A_R^4}+\frac{5(1-2\omega_R^2)}{A_R^3}+\frac{2\omega_R(3-4\omega_R^2)}{A_R^2}\nonumber\\
&+&\frac{4(2\omega_R^4+2\omega_R^2-1)}{A_R}+20\omega_R (2\omega_R^2-1)\nonumber\\
&+&12 (2\omega_R^2-1)A_R\,,\nonumber\\
c&=& -\frac{47}{3A_R^3}-\frac{32\omega_R}{A_R^2}+\frac{3+16\omega_R^2-4\omega_R^4}{A_R}\nonumber\\
&+&20\omega_R (2\omega_R^2+3)+4(3+26\omega_R^2)A_R+48\omega_R A_R^2\,.\nonumber\\
\eea
Here,  the various retarded scalar quantities $\rho_R$, $\omega_R$, $A_R$ as well as the retarded vectors $A_R^\alpha$, $v_R^\alpha$ are defined in Appendix \ref{App:retarded} (in the mostly plus signature), together with their advanced counterparts~\footnote{This notation is adapted from Ref. \cite{Bel:1981be}, where the two particles (and related particle-dependent objects) were denoted as $m$ and $m'$, instead of $m_1$ and $m_1$.}.
In the expression of $\Gamma_{1   \rm R\,, 1PM}^\alpha$ one has considered that the world lines were curved, and satisfied their equations of motion, at least at the required accuracy. In our present 2PM-accurate setting this means that the retarded quantities $z_{2}^\mu[\tau_{2R}(\tau_1)]$ and $u_{2\rm R}^\mu(\tau_1)=u_2^\mu(\tau_2)\big|_{\tau_2=\tau_{2R}(\tau_1)}$ entering $\Gamma_{1   \rm R\,, 1PM}^\alpha(\tau_1)$ must be computed along world lines satisfying the 1PM equations of motion.
By contrast, the evaluation of $\Gamma_{1   \rm R\,, 2PM}^\alpha$ can be done in the leading-order (LO) approximation  where the curvature of the world lines is neglected.

In the (past-asymptotic) flat spacetime a basis of vectors (denoted by a bar) is naturally associated with the asymptotic incoming four-velocities of the two bodies, $\bar u_1$ and $\bar u_2$, together with two additional 
``initial" positions, $z_1^\mu(0),z_2^\mu(0)$, taken at $\tau_a=0$. The corresponding world lines have parametric  equations of the form
\bea
z_{1}^\mu(\tau_1) &=&z_1^\mu(0)+\tau_1 \bar u_{1}^\mu+ \delta^G  z_{1}^\mu(\tau_1) + O(G^2)\,,\nonumber\\
z_{2}^\mu(\tau_2)&=&z_2^\mu(0)+\tau_2 \bar u_{2}^\mu+ \delta^G  z_{2}^\mu(\tau_1) + O(G^2)\,.
\eea

Here $z_a^\mu(0)$ (corresponding to $\tau_a=0$) denotes the \lq\lq midpoint" on each world line around which the 
magnitude of the 1PM acceleration is time-symmetric.~\footnote{As discussed below, at order $G$, the {\it retarded} 1PM acceleration is equal to the advanced one and it is time-symmetric.} The $O(G)$ correction 
$\delta^G  z_{a}^\mu(\tau_a)$ (written below) is normalized so as to vanish at  $\tau_a=0$. [One cannot
impose the boundary condition that $\delta^G  z_{a}^\mu(\tau_a)$ vanishes at $\tau_a \to -\infty$ because
of the logarithmic divergence of the world lines away from straight world lines in the incoming (and outgoing)
states.]

The explicit expressions of the (retarded) 1PM-level terms $\delta^G  z_{a}^\mu(\tau_a)$ and
$\delta^G  u_{a}^\mu(\tau_a)= \frac{d}{d \tau_a} \delta^G  z_{a}^\mu(\tau_a)$ read, for $a=1$
(see Eqs. (4.4) and (4.5) of Ref. \cite{Bini:2018ywr}),
\begin{widetext}
\bea
\label{OG_corr_wln}
\delta^G z_1^\mu (\tau_1)
&=&  +Gm_2 (1-2\gamma^2) \frac{(S(\tau_1)-1)}{(\gamma^2-1)}\, \frac{b_{12}^\mu}{b_0} 
+Gm_2\frac{\gamma (2\gamma^2-3) }{ (\gamma^2-1)^{3/2} }  
\ln \left( S(\tau_1)\right)\, (\bar u_2^\mu -\gamma \bar u_1^\mu)\,,\qquad \nonumber\\
\delta^G u_1^\alpha (\tau_1)  
&=&  +G m_2  \frac{(1-2\gamma^2)S(\tau_1)}{\sqrt{\gamma^2-1}D(\tau_1)}\, \frac{b_{12}^\mu}{b_0}    
+Gm_2 \frac{\gamma (2\gamma^2-3)}{(\gamma^2-1)D(\tau_1)} (\bar u_2^\mu -\gamma \bar u_1^\mu)\,,
\end{eqnarray}
\end{widetext}
with analog expressions for $a=2$ obtained by exchanging $1\leftrightarrow 2$.
Here, $\gamma\equiv -\bar u_1 \cdot \bar u_2$ is the Lorentz factor between the two incoming world lines and $b_{12}^\mu$ is a 1PM-accurate vectorial (spatial) impact parameter. More precisely, it connects the midpoints of the two world lines and its magnitude, $b_0$, measures the closest approach distance,   
\beq
 b_{12}^\mu=z_{1}^\mu(0)- z_{2}^\mu(0) \,\,  ; \,\, b_0= |b_{12}^\mu|=|z_{1}^\mu(0)- z_{2}^\mu(0)|\,.
\eeq
The vectorial impact parameter $b_{12}^\mu$ has been  chosen to be orthogonal to the two incoming four velocities $\bar u_1$ and $\bar u_2$. The auxiliary functions $S(\tau)$ and $D(\tau)$ entering Eq. \eqref{OG_corr_wln} are defined as follows
\bea
\label{D_and_S_def}
D(\tau)&=&\sqrt{b_0^2 +\tau^2 (\gamma^2-1)}\,, \nonumber\\
S(\tau)&=&\frac{1}{b_0}\left(\tau \sqrt{\gamma^2-1} +D(\tau)\right)\,,
\eea
so that $D(0)=b_0$ and $S(0)=1$, thereby ensuring that $\delta^G z_a^\mu (\tau_a=0)=0$. [By contrast, $\delta^G u_a^\alpha (\tau_1)$ does not vanish at $\tau_a=0$ but  
vanishes at $\tau_a\to -\infty$  
because of our chosen boundary conditions.] In addition, we have the identity $S(\tau)S(-\tau)=1$.

When working in the (incoming) rest frame of particle 1, $e_0^\mu=\bar u_1^\mu$, as we shall often do below, 
it is  convenient to introduce two spatial unit vectors
\bea
e_x^\mu&=&\hat  b_{12}^\mu=\frac{ b_{12}^\mu}{b_0}\,,\nonumber\\
e_y^\mu&=& -\frac{\bar u_2^\mu-\gamma \bar u_1^\mu}{\sqrt{\gamma^2-1}}\,.
\eea
Note that $e_y$ is linked to the projection of $\bar u_2$ orthogonally to $u_1$ in the following way
\bea 
v&=&P(\bar u_1)\bar u_2 \equiv \bar u_{2\perp 1}=\bar u_2-\gamma \bar u_1\nonumber\\
&=&-\sqrt{\gamma^2-1}e_y\,,
\eea
with the projector $P(u)$ orthogonal to the timelike direction $u$ ($u\cdot u=-1$) defined as
\beq
P(u)^\mu{}_\nu=\delta^\mu{}_\nu+u^\mu u_\nu\,.
\eeq

The third spatial vector ${\bf e}_z={\bf e}_x\times {\bf e}_y$ does not enter the parametrization of the two world lines because the motion takes place in   the $x$-$y$ plane.
The minus sign in the definition of $e_y^\mu$ has been chosen so that  the center-of-mass (c.m.) angular momentum is aligned with the $z$-axis: $J^{xy}_{\rm c.m.}=J^z_{\rm c.m.}>0$.
At the 1PM approximation the magnitude of the c.m. angular momentum is (see Appendix D of Ref. \cite{Bini:2018ywr})
\beq
\label{Jcm}
J_{\rm c.m.}=b\, P_{\rm c.m.}\,,  
\eeq
where $P_{\rm c.m.}$ is the common magnitude of the two incoming spatial linear momenta in the c.m. frame, and where the  (incoming) impact parameter $b_{\rm in}$ 
is given by
\beq
\label{impact_eq}
b_{\rm in}=b_0 +G(m_1+m_2)\frac{(2\gamma^2-1)}{\gamma^2-1}+O(G^2)\,.
\eeq
Note that $b_{\rm in}$  differs from the  minimal approaching distance $ b_0= |b_{12}^\mu|=|z_{1}^\mu(0)- z_{2}^\mu(0)|$ by terms of order $G$.

In the incoming rest frame of particle 1 defined above the explicit expressions of the incoming four velocities read
\bea
\bar u_{1}^\mu&=&\delta_0^\mu\,,\qquad
\bar u_{2}^\mu=\gamma \delta_0^\mu -\sqrt{\gamma^2-1}\delta_y^\mu\,,
\eea
while the explicit expressions of the two world lines read  (if one takes $z_1^\mu(0)$ and $z_2^\mu(0)$
in the  form  $z_a^\mu(0)= b_a e_x^\mu$) 
\begin{widetext}
\bea
\label{OG_corr_wl}
z_1^\alpha (\tau_1) 
&=&b_1 e^\alpha_x+ \bar u_1^\alpha \tau_1  +Gm_2 (1-2\gamma^2) \frac{(S(\tau_1)-1)}{(\gamma^2-1)}\, e^\alpha_x 
-Gm_2\frac{\gamma (2\gamma^2-3) }{ \gamma^2-1 }  
\ln \left( S(\tau_1)\right)\,  e^\alpha_y\,,\nonumber\\
u_1^\alpha (\tau_1)  
&=&  \bar u_1^\alpha +G m_2  \frac{(1-2\gamma^2)S(\tau_1)}{\sqrt{\gamma^2-1}D(\tau_1)}\, e^\alpha_x   
-Gm_2 \frac{\gamma (2\gamma^2-3)}{\sqrt{\gamma^2-1}D(\tau_1)}  e^\alpha_y\,,\nonumber\\
z_2^\alpha(\tau_2)  &=&b_2 e^\alpha_x+ \bar u_2 ^\alpha \tau_2  -Gm_1(1-2\gamma^2)\frac{(S(\tau_2)-1)}{(\gamma^2-1)}\, e^\alpha_x  
+Gm_1 \frac{\gamma (2\gamma^2-3)}{(\gamma^2-1)^{3/2}}  
\ln \left(S(\tau_2) \right)\, v'{}^\alpha  \nonumber\\
u_2^\alpha(\tau_2)&=& \bar u_2^\alpha   -G m_1 \frac{(1-2\gamma^2)S(\tau_2)}{\sqrt{\gamma^2-1}D(\tau_2)}\,  e^\alpha_x    
+Gm_1  \frac{\gamma(2\gamma^2-3)}{(\gamma^2-1)D(\tau_2)}  v'{}^\alpha 
\,.
\end{eqnarray} 
When performing explicit calculations it is useful to choose $b_1=b_0>0$ and $b_2=0$.
\end{widetext}
Here, we have used the notation (see Ref. \cite{Bini:2021gat},  Eq. (3.46) there)
\bea 
v'&=&P(\bar u_2)\bar u_1\equiv \bar u_{1\perp 2} =\bar u_1-\gamma \bar u_2\nonumber\\
&=&-(\gamma^2-1)e_0 +\gamma \sqrt{\gamma^2-1}e_y\,.
\eea

Inserting in $\Gamma_{a   \rm R\,, 1PM}^\alpha$ the explicit expressions of the 1PM-accurate world lines 
(as functions of $\tau_a$) re-defines   $F_{a\rm R}^\mu$ as a function of $\tau_a$ (rather than as a functional of the world lines) that we will denote as ${\mathcal F}_{a\rm R}^\mu(\tau_a)$ to distinguish it from the original world line-dependent force, Eq. \eqref{wl_dep_force}. Expanding ${\mathcal F}_{a\rm R}^\mu(\tau_a)$ in powers of $G$ then yields 
\beq
\label{master_eq2}
{\mathcal F}_{a\rm R}^\mu(\tau_a)= {\mathcal F}_{a\rm R}^{G\,\mu}(\tau_a)+{\mathcal F}_{a\rm R}^{G^2\,\mu}(\tau_a)+O(G^3)\,.
\eeq
The 1PM expression of the force, ${\mathcal F}_{a\rm R}^{G\,\mu}(\tau_a)$, is obtained by inserting the LO straight line world lines, namely
\beq
\bar z_a(\tau_a)= z_a(0)+\bar u_a \tau_a\,,
\eeq
in $F_{a\rm R\,, 1PM}^\mu$ and explicitly reads
\bea
\label{F_with_perp}
{\mathcal F}_{a\rm R}^{G\,\mu}(\tau_a)&=&\frac{Gm_a m_b}{|\bar z_a(\tau_a)-\bar z_{b\perp}(\tau_a)|^3}\left[4(\gamma^2-1)b_{ab}^\mu\right.\nonumber\\
&&\left.-(2\gamma^2-3)(\delta^\mu_\nu+ \bar u_a^\mu \bar u_{a\nu})(\bar z_a^\nu(\tau_a)-\bar z_{b\perp}^\nu(\tau_a) )\right]\,,\nonumber\\
\eea 
where $a=1,2$, $b\not =a$, $b_{ab}^\mu=z_a^\mu(0)-z_b^\mu(0)$,  and  $\bar z_{b\perp}^\nu(\tau_a)$ is the foot of the
perpendicular  of the point $\bar z_a^\nu(\tau_a)$ on the (straight) line ${\mathcal L}_b$, so that
\bea
\label{z_2_perp_definition}
\bar z_a^\nu(\tau_a)-\bar z_{b\perp}^\nu(\tau_a)&=&\bar z_a^\nu(\tau_a)-  z_{b}^\nu(0)\nonumber\\
&+&[\bar u_b\cdot (\bar z_a(\tau_a)- z_b(0)] \bar u_b^\nu\,.\qquad
\eea 
Explicitly, for $a=1$,  we have $|\bar z_1(\tau_1)-\bar z_{2\perp}(\tau_1)|=D(\tau_1)$ and
\bea
\label{F_with_perp2}
{\mathcal F}_{1\rm R}^{G\,\mu}(\tau_1)&=&\frac{Gm_1 m_2}{D(\tau_1)^3}\left[(1-2\gamma^2)b_{12}^\mu\right. \nonumber\\
&&\left.+ \tau_1 (2\gamma^2-3)\gamma(\gamma \bar u_1^\mu-\bar u_2^\mu) \right]\,.
\eea 

By contrast, ${\mathcal F}_{a\rm R}^{G^2\,\mu}(\tau_a)$ is obtained as the sum of two contributions, one coming from inserting  in $F_{a\rm R\,, 1PM}^\mu$ the $O(G)$-corrected  world lines, Eq. \eqref{OG_corr_wl}, and the other one coming directly from $F_{a\rm R\,, 2PM}^\mu$.
We will not need here the (complicated) explicit expression of the retarded second-order force ${\mathcal F}_{a\rm R}^{G^2\,\mu}$, but only of its time-odd part, see below.

\section{Retarded vs advanced force at $O(G^2)$}

In the previous section we considered the physical, {\it retarded} equations of motions of two gravitationally interacting point masses. In order to decompose this retarded dynamics in a conservative part and a dissipative part it is useful, in our present PM framework\footnote{When working in a Post-Newtonian (PN) framework, an alternative way to see the presence of dissipative effects is to expand in powers of $v/c$, and to extract its odd part under velocity-reversal, as was done in Ref. \cite{Damour:1981bh}.},  to consider the advanced counterpart of the equations of motion.

Introducing an indicator $\epsilon$, with $\epsilon=1$ in the retarded case and $\epsilon=-1$ in the advanced one,  the generalized equations of motion are obtained as follows.  

The generalized form (valid for $\epsilon=\pm 1$) of the world-line-functional version of the equations of motion reads
\beq
m_a\frac{d^2z_a^\mu(\tau_a)}{d\tau_a^2} =F_{a\epsilon}^\mu[z_a(\tau_a),u_a(\tau_a); z_{b\epsilon}(\tau_a), u_{b\epsilon}(\tau_a)] \,.
\eeq
This then yields the corresponding  $\tau_a$-dependent generalized forces
\beq
{\mathcal F}_{a\epsilon}^\mu(\tau_a)= {\mathcal F}_{a\epsilon}^{G\,\mu}(\tau_a)+{\mathcal F}_{a\epsilon}^{G^2\,\mu}(\tau_a)+O(G^3)\,.
\eeq

As is evident from Eqs. \eqref{F_with_perp}-\eqref{F_with_perp2}, the $O(G)$ force in this equation, ${\mathcal F}_{a\epsilon}^{G\,\mu}(\tau_a)$, is time-symmetric, and does not depend on $\epsilon$ (we henceforth denote it with a label \lq\lq$0$" in place of $\epsilon$),
\beq
\label{F_a_0}
{\mathcal F}_{a\epsilon}^{G\,\mu}(\tau_a)={\mathcal F}_{a 0}^{G\,\mu}(\tau_a)\,.
\eeq
The time-asymmetry in the $\tau_a$-dependent version of the equations of motion only enters at order $G^2$, as is discussed in detail below. 

The $\epsilon$-dependent  world line equations of motion read
\beq
\frac{du_{a\epsilon}^\alpha}{d\tau_a}=\Gamma_{a\, \epsilon ,\rm 1PM}^\alpha(\tau_a)+\Gamma_{a\, \epsilon ,\rm 2PM}^\alpha(\tau_a)+O(G^3)\,.
\eeq
For $a=1$, one has
\bea
\label{eps_eqs}
\Gamma_{1\, \epsilon , \rm 1PM}^\alpha&=&\frac{Gm_2}{\rho_\epsilon^2}[(1-2\omega_\epsilon^2)A_\epsilon^\alpha \nonumber\\
&-&\epsilon (1+2\omega_\epsilon^2 +4A_\epsilon \omega_\epsilon )v_\epsilon^\alpha]\,, \nonumber\\
\Gamma_{1\, \epsilon , \rm 2PM}^\alpha  
&=& \frac{G^2}{\rho_\epsilon^3}[m_2^2 (a_0 A_\epsilon ^\alpha +\epsilon c_0 v_\epsilon^\alpha) \nonumber\\
&+& m_1m_2 (aA_\epsilon^\alpha +c \epsilon v_\epsilon^\alpha)]\nonumber\\ 
&-& 4m_1\epsilon  \frac{d}{d\tau_1 } [\Gamma_{1\,\epsilon 1PM}^\alpha \ln A_\epsilon ]\,.
\eea
The explicit expressions of the $\epsilon$-dependent quantities entering these equations are defined in Appendix \ref{App:retarded}. [The formal expression of the coefficients $a_0, c_0$ and $a, c$ are obtained from those listed above in Eqs. \eqref{ac_coeffs} by replacing $R$ by $\epsilon$.] Let us also display  the following intermediate results (where $\delta\equiv \delta^{G}$ and $O(G^2)$ error terms are implicit)
\bea
\tau_{2\epsilon}(\tau_1)&=& \gamma \tau_1 - \epsilon D(\tau_1)+ \delta \tau_{2\epsilon}(\tau_1)\,,\nonumber\\
\rho_\epsilon(\tau_1)&=&D(\tau_1)+ \delta \rho_\epsilon(\tau_1)\,,\nonumber\\
\omega_\epsilon(\tau_1)&=&-\gamma + \delta \omega_\epsilon(\tau_1)\,, \nonumber\\
A_\epsilon(\tau_1)&=&\gamma -\epsilon \frac{\gamma^2-1}{D(\tau_1)}\tau_1+ \delta A_\epsilon(\tau_1)\,,\nonumber\\
A^\mu_\epsilon(\tau_1)&=&\frac{1}{D(\tau_1)}\left[b_0e_x^\mu +\sqrt{\gamma^2-1}(\gamma \tau_1 - \epsilon D(\tau_1))e_y^\mu\right]\nonumber\\
&+&  \delta A^\mu_\epsilon(\tau_1)\,,\nonumber\\
v^\mu_\epsilon(\tau_1)&=&-\sqrt{\gamma^2-1}e_y^\mu+ \delta v^\mu_\epsilon(\tau_1)\,,
\eea
with $O(G)$-corrections given by  
\bea
\delta \tau _{2\epsilon}(\tau_1)&=&\frac{1}{D}\left[\frac{\sqrt{\gamma^2-1}}{\gamma} \epsilon\tau_1  \delta z_2^y(\tau_{2\epsilon}(\tau_1))\right. \nonumber\\
&-&\epsilon\sqrt{\gamma^2-1}(\gamma \tau_1-\epsilon D(\tau_1)) \delta z_1^y(\tau_1)\nonumber\\
&+& \left.\epsilon b_0\delta z_2^x(\tau_{2\epsilon}(\tau_1))
-\epsilon b_0\delta z_1^x(\tau_1)
\right]\,,\nonumber\\
\delta \omega_\epsilon(\tau_1)&=&  \frac{\sqrt{\gamma^2-1}}{\gamma} \delta u_2^y(\tau_{2\epsilon}(\tau_1))-\delta u_1^y(\tau_1) \sqrt{\gamma^2-1}\,,\nonumber\\
\delta \rho_\epsilon(\tau_1)&=& -\epsilon b_0 \delta u_2^x(\tau_{2\epsilon}(\tau_1))
-\epsilon \frac{\sqrt{\gamma^2-1}}{\gamma} \tau_1\delta u_2^y(\tau_{2\epsilon}(\tau_1))\nonumber\\
&-&\frac{b_0}{D(\tau_1)}  \delta z_2^x(\tau_{2\epsilon}(\tau_1))+\frac{b_0}{D(\tau_1)}  \delta z_1^x(\tau_1)\nonumber\\
&-&\tau_1 \frac{\sqrt{\gamma^2-1}}{\gamma D(\tau_1)}\delta z_2^y(\tau_{2\epsilon}(\tau_1))\nonumber\\
&+&\frac{\gamma}{D(\tau_1)}\sqrt{\gamma^2-1} \tau_1  \delta z_1^y(\tau_1)\,,\nonumber\\
\delta v^\mu_\epsilon(\tau_1)&=& 
-\delta u_1^y(\tau_1)\sqrt{\gamma^2-1}e_0^\mu  \nonumber\\
&+& \left(\delta u_2^x(\tau_{2\epsilon}(\tau_1))  -\delta u_1^x(\tau_1)\gamma\right)e_x^\mu \nonumber\\
&+& \left(\delta u_2^y(\tau_{2\epsilon}(\tau_1)) -\delta u_1^y(\tau_1)\gamma \right)e_y^\mu\,,  
\eea
and similarly for $\delta A^\mu_\epsilon(\tau_1)$ and $\delta A_\epsilon(\tau_1$), not shown here because involving longer expressions. All quantities here are supposed to be functions of $\tau_1$.
In view of Eq. \eqref{F_a_0}, the 1PM-accurate solutions of the $\epsilon$-dependent equations of motion, Eqs.  \eqref{eps_eqs}, coincide with the retarded solution displayed [as functional of $\bar u_a$ and $z_a(0)$] in Eq.\eqref{OG_corr_wl} above, 
\bea
z_{a\epsilon}^\alpha(\tau_a)&=&z_{aR}^\alpha(\tau_a) +O(G^2)\,,\nonumber\\
u_{a\epsilon}^\alpha(\tau_a)&=&u_{aR}^\alpha(\tau_a)+O(G^2)\,.
\eea

\section{Time-symmetric dynamics at $O(G^2)$}
\label{time_sym_dyn}

Our aim here is to decompose the retarded two-body dynamics in conservative and dissipative parts. To do this
in a PM framework, one can
first define a time-symmetric version of the 2PM dynamics by solving Einstein's equations from the start with a time-symmetric Green's function in 4 dimensions,
\beq
G_{\mu\nu\alpha\beta}^{\rm S}(x,y)= P_{\mu\nu\alpha\beta}\delta ((x-y)^2)\,,
\eeq
where $P_{\mu\nu\alpha\beta}=\eta_{\mu\alpha}\eta_{\nu\beta}-\frac1{2}\eta_{\mu\nu}\eta_{\alpha\beta}$. It can be easily checked that, at order $G^2$ included,  the use of such a time-symmetric propagator leads to equations of motion involving the following time-symmetric forces
\bea \label{FSvsRA}
F_{a S}^\mu[z_{a},u_{a}, z_{bR}, u_{bR}, z_{bA}, u_{bA}]&=&\frac12 (F_{a R}^\mu+F_{a A}^\mu)\nonumber\\
&+& O(G^3)\,,
\eea
with corresponding $\tau_a$-dependent forces
\bea \label{calFSvsRA}
{\mathcal F}_{aS}^\mu(\tau_a)&=&{\mathcal F}_{a  0}^{G\,\mu}(\tau_a)+\frac12 ({\mathcal F}_{a\rm R}^{G^2\,\mu}(\tau_a)+{\mathcal F}_{a\rm A}^{G^2\,\mu}(\tau_a))\nonumber\\
&+&O(G^3)\,.
\eea 

When using the time-symmetric Green's function, the dynamics can be derived from a Fokker(-Wheeler-Feynman) action \cite{Fokker:1929,Wheeler:1949hn,Friedman:2005rx}
 of the general form\footnote{In some of the following expressions $\tau_a$ are unnormalized, generic worldline
 parameters.}
\bea
I&=&-\sum_{a} m_a \int d\tau_a (-\dot z_{a\mu} \dot z_a^{\mu})^{1/2}\nonumber\\
&+&\sum_{a<b}\iint 
d\tau_a d\tau_b \Lambda_{ab}+\sum_{a<b<c}\iiint d\tau_a d\tau_bd\tau_c \Lambda_{abc}\nonumber\\
& +& O(G^3)\,,
\eea 
where the one-graviton exchange ($O(G)$) contribution explicitly  reads (see Eq. (22) of Ref. \cite{Friedman:2005rx})
\bea
\label{Lambda_ab_def}
&&\Lambda_{ab} (z_a-z_b, \dot z_a, \dot z_b)=2G m_am_b \delta [(z_a-z_b)^2]  \times \qquad\qquad\nonumber\\
&&\qquad \frac{(\dot z_a\cdot \dot z_b)^2-\frac12 (\dot z_a\cdot \dot z_a)(\dot z_b\cdot \dot z_b) }{(-\dot z_a\cdot \dot z_a)^{1/2}(-\dot z_b\cdot \dot z_b)^{1/2}}\,,\qquad\nonumber\\
&&\qquad =2G m_am_b \delta [(z_a-z_b)^2]P_{\mu\nu\alpha\beta}u_a^\mu u_a^\nu u_b^\alpha u_b^\beta\,,
\eea
and where $\Lambda_{abc}$ refers to the one-loop, order $G^2$, interaction, etc.

The existence of such a Poincar\'e-invariant Fokker action guarantees that the corresponding time-symmetric dynamics has all the Noetherian conserved quantities associated with the Poincar\'e symmetries, namely total linear momentum and total angular momentum of the system \cite{Wheeler:1949hn,Dettman:1954zz,Friedman:2005rx} 
\bea
\label{P_and_J}
P_\mu^{\rm sys}(\tau_1,\tau_2)&=& P_{\mu}^{\rm kin}(\tau_1,\tau_2)   +  P_{\mu}^{\rm int}(\tau_1,\tau_2)\,, \nonumber\\
J_{\mu\nu}^{\rm sys}(\tau_1,\tau_2)&=&J_{\mu\nu}^{\rm kin}(\tau_1,\tau_2)+J_{\mu\nu}^{\rm int}(\tau_1,\tau_2)\,.
\eea
The conserved quantities of the system are obtained as the sum of kinematical contributions ($ P_{\mu}^{\rm kin}(\tau_1,\tau_2)$, $J_{\mu\nu}^{\rm kin}(\tau_1,\tau_2)$)  and (field-mediated) interaction ones ($P_{\mu}^{\rm int}(\tau_1,\tau_2)$, $J_{\mu\nu}^{\rm int}(\tau_1,\tau_2)$).
The kinematical contributions read
\bea
\label{P_and_J_kin}
P_{\mu}^{\rm kin}(\tau_1,\tau_2)&=&\sum_a m_a  u_{a\mu}(\tau_a)\,,\nonumber\\
J_{\mu\nu}^{\rm kin}(\tau_1,\tau_2)&=& \sum_a m_a  (z_a(\tau_a)  \wedge  u_a(\tau_a))_{\mu\nu}\,,
\eea
where the wedge product symbol is  defined as
\beq
(A\wedge B)^{\mu\nu}\equiv A^\mu B^\nu-A^\nu B^\mu\equiv A^\mu \wedge B^\nu\,.
\eeq 
The  \lq\lq interaction" or \lq\lq Fokker" parts of the linear momentum and of the angular momentum read, at 
the one-graviton exchange level,
\bea
\label{P_and_J_field}
P^{\rm int}_{\alpha}(\tau_1,\tau_2)&=& \int_{-\infty}^\infty d\bar \tau \frac{\partial \Lambda}{\partial \dot z_1^\alpha}(\tau_1,\bar \tau)
+\int_{-\infty}^\infty d \tau \frac{\partial \Lambda}{\partial \dot z_2^\alpha}(\tau , \tau_2)\nonumber\\
&+& 2\left(\int_{\tau_1}^{\infty}\int_{-\infty}^{\tau_2}
-\int_{-\infty}^{\tau_1}\int_{\tau_2}^{\infty}\right)d\tau d\bar\tau \times \nonumber\\
&& (z_1(\tau)-z_2(\bar\tau))_\alpha\frac{\partial \Lambda}{\partial w}\,,\nonumber\\
J_{\mu\nu}^{\rm int}(\tau_1,\tau_2)&=&\int_{-\infty}^\infty  d\bar \tau [(z_1\wedge {\mathcal P}_1)(\tau_1,\bar\tau)]_{\mu\nu}\nonumber\\
&+&\int_{-\infty}^\infty  d\tau [(z_2\wedge {\mathcal P}_2)(\tau,\tau_2)]_{\mu\nu}\nonumber\\
&+&\left(\int_{\tau_1}^{\infty}\int_{-\infty}^{\tau_2}
-\int_{-\infty}^{\tau_1}\int_{\tau_2}^{\infty}\right)
\left[
(z_1\wedge {\mathcal Q})(\tau,\bar\tau)\right.\nonumber\\
&+&\left.(\dot z_1\wedge {\mathcal P}_1)(\tau,\bar\tau)
\right]_{\mu\nu}
d\tau d\bar \tau\,.
\eea
Considering $\Lambda=\Lambda_{12}$ (see Eq. \eqref{Lambda_ab_def} above) as a function of 
\beq
w\equiv R_\beta R^\beta=(z_1-z_2)^2\,,\quad
 R_\beta=(z_1-z_2)_\beta\,,
\eeq
as well as of  $\dot z_1^\mu$, $\dot z_2^\mu$, $\Lambda=\Lambda(w,\dot z_1^\mu,\dot z_2^\mu)$,
the values of the quantities entering the interaction terms are
\bea
{\mathcal P}_{1\beta}&=&\frac{\partial \Lambda}{\partial \dot z_1^ \beta  }\,,\nonumber\\ 
{\mathcal P}_{2\beta}&=&\frac{\partial\Lambda}{\partial\dot {z_2}{}^\beta} \,,\nonumber\\
{\mathcal Q}_\beta&=&\frac{\partial \Lambda}{\partial R^\beta} 
=2(z_1-z_2)_\beta \frac{\partial \Lambda}{\partial w}\,.
\eea 
At order $G$ the interaction terms only depend on two finite segments on the world lines.
This fact means, in particular, that there are no logarithmic divergences (and related logarithmic ambiguities) in the definition of $J^{\mu\nu}_{\rm sys}$.

The  conservation of the system Noetherian quantities  means their independence on $\tau_1$ and $\tau_2$. 
In the present work we will not need the explicit expressions of the total Noetherian quantities at order $G^2$. What is important for us is only the fact that they exist for the time-symmetric dynamics. We will, however, present an explicit computation of the Noetherian quantities at the 1PM accuracy in order to see the importance of the presence of interaction terms to ensure the conservation of manifestly Poincar\'e-invariant total linear momentum and angular momentum.

We will explicitly compute below at order $G$ the Noetherian  quantities $P_\mu^{\rm sys}$ and $J_{\mu\nu}^{\rm sys}$  and show their conservation.
The existence of interaction contributions (starting at the $O(G)$ level) to both linear momentum and angular moment is well-known in the PN context.
For instance, at the 1PN level (see Ref. \cite{[23]81} for the 2PN case), the Poincar\'e-invariance of the (harmonic coordinates) Lorentz-Droste-Einstein-Infeld-Hoffmann Lagrangian yields both  a conserved 1PN-accurate linear momentum $P^\mu=((m_1+m_2)c^2+E, P^i)$ and a 1PN-accurate angular momentum $J^{\mu\nu}$.
Explicitly (with $v_a^i= d z_a^i/dt$)
\bea
E&=&\frac12 m_1 v_1^2 -\frac{Gm_1m_2}{2r_{12}}+\frac{1}{c^2}\left[\frac38 m_1 v_1^4 \right.\nonumber\\
&+&\frac{Gm_1m_2}{r_{12}}\left(-\frac14 (n_{12}v_1)(n_{12}v_2)+\frac32 v_1^2-\frac74 (v_1v_2)\right) \nonumber\\
&+&\left. \frac{G^2m_1^2 m_2}{2r_{12}^2}\right]+1\leftrightarrow 2\,,
\eea
and
\beq
P^i=p_1^i+p_2^i\,,
\eeq
where 
\bea
{\mathbf p}_1&=&m_1{\mathbf v}_1+\frac{1}{c^2} \left[
\frac12 m_1 v_1^2 {\mathbf v}_1\right.\nonumber\\
&+&\left.\frac{Gm_1m_2}{r_{12}}\left(-\frac12 (n_{12}v_2){\mathbf n}_{12}+3{\mathbf v}_1-\frac72 {\mathbf v}_2\right)\right]\nonumber\\
\eea
with ${\mathbf p}_2$ obtained by exchanging $1\leftrightarrow 2$.

The spatial components of the  
angular momentum $J_{ij}=\epsilon_{ijk}J_k$ are 
\bea
{\mathbf J}&=& {\mathbf z}_1\times {\mathbf p}_1+{\mathbf z}_2\times {\mathbf p}_2\nonumber\\
&=& M_1 {\mathbf z}_1\times {\mathbf v}_1+M_2 {\mathbf z}_2\times {\mathbf v}_2\nonumber\\
&+&\frac{1}{c^2} \frac{Gm_1m_2}{r_{12}}((n_{12}v_1)+(n_{12}v_2)){\mathbf z}_1\times {\mathbf z}_2\,,\qquad
\eea
where
\beq
M_a=m_a+\frac{1}{c^2}\left(\frac12 m_a v_a^2 -\frac12 \frac{Gm_am_b}{r_{12}}\right)\,,
\eeq
while the  conserved boost generator reads
\beq
K^i=J^{i0}=M_1 z_{1}^i+M_2 z_{2}^i -t (p_1^i+p_2^i)\,.
\eeq

\section{Radiation reaction at $O(G^2)$}

Coming back to  the physical, retarded dynamics, the results, Eqs. \eqref{FSvsRA}, \eqref{calFSvsRA}, show 
 that one can decompose the $\tau_a$-dependent force entering the retarded dynamics
  in conservative (time-symmetric) and radiation-reaction (time-antisymmetric) parts as follows
\beq
\label{deco_F}
{\mathcal F}_{aR}^\mu(\tau_a)={\mathcal F}_{aS}^\mu(\tau_a)+{\mathcal F}_{a\,\rm rr}^\mu(\tau_a)\,,
\eeq
where the radiation-reaction part of the force reads
\beq
\label{def_f_rr_fin}
{\mathcal F}_{a\,\rm rr}^\mu(\tau_a)=\frac{1}{2}({\mathcal F}_{aR}^\mu(\tau_a)-{\mathcal F}_{aA}^\mu(\tau_a))+O(G^3)\,.
\eeq

\subsection{Explicit expressions of the $G^2$-accurate radiation-reaction force}
\label{exp_expr_g2rr}

We explicitly computed the $O(G^2)$ expressions of the radiation-reaction force ${\mathcal F}_{a\,\rm rr}^\mu(\tau_a)$. In the rest frame of particle 1 it has only two nonzero components, 
${\mathcal F}_{1\,\rm  rr}^x(\tau_1)$ and ${\mathcal F}_{1\,\rm  rr}^y(\tau_1)$:
\bea
{\mathcal F}_{1\,\rm rr}^\mu(\tau_1)&=&{\mathcal F}_{1\,\rm  rr}^x(\tau_1) e_x^\mu+{\mathcal F}_{1\,\rm  rr}^y(\tau_1) e_y^\mu\nonumber\\
&=& {\mathcal F}_{1\,\rm  rr}^x(\tau_1) \hat b_{12}^\mu -\frac{{\mathcal F}_{1\,\rm  rr}^{y}(\tau_1)}{\sqrt{\gamma^2-1}} (\bar u_2^\mu-\gamma \bar u_1^\mu)\,.\qquad
\eea
In order to write the explicit expressions of the $x$ and $y$ components of ${\mathcal F}_{1\,\rm  rr}^\mu(\tau_1)$ is convenient to rescale ${\mathcal F}_{1\,\rm  rr}^\mu(\tau_1)$, namely
\bea
{\mathcal F}_{1\,\rm  rr}^\mu(\tau_1) &=& \frac{G^2m_1^2m_2  }{D^5(\tau_1)} \,\tilde F^\mu(\tau_1)\,,
\eea
where
\beq
\label{lo_no-log}
\tilde F^\mu(\tau_1)=\tilde F^\mu_{\rm log}(\tau_1)  +  \tilde F^\mu_{\rm no-log}(\tau_1) \,. 
\eeq
Moreover, $D(\tau_1)$ and ${\mathcal D}(\tau_1)$ (which enters Eq.\eqref{D_and_cal_D_defs} below) are the positive roots of
\bea
\label{D_and_cal_D_defs}
D^2(\tau_1)&=& b_0^2+p_\infty^2 \tau_1^2\,,\nonumber\\
{\mathcal D}^2(\tau_1)
&=&D^2(\tau_1)+p_\infty^2 b_0^2= \gamma^2 \left( b_0^2 + v^2 \tau_1^2  \right) \,,
\eea
where
\beq
\label{v_def}
v \equiv\sqrt{1-\frac{1}{\gamma^2}}=\frac{p_\infty}{\sqrt{1+p_\infty^2}}\,,
\eeq
denotes the relative velocity between the incoming particles.

The log-part of $\tilde F^\mu(\tau_1)$, Eq. \eqref{lo_no-log}, can be written as
\beq
\tilde F^\mu_{\rm log}(\tau_1)=B_1^\mu \ln \left(\frac{{\mathcal D}^2}{D^2}\right)+B_2^\mu {\mathcal A}(p_\infty)
\eeq
for both components, 
with
\beq
\label{mathcal_A}
{\mathcal A}(p_\infty)=\frac12 \ln \left( \frac{\sqrt{1+p_\infty^2}+p_\infty}{\sqrt{1+p_\infty^2}-p_\infty} \right)\,,
\eeq
and
\bea
B_1^x&=& -6p_\infty^2 (2p_\infty^2+1) b_0 \tau_1\,, \nonumber\\
B_2^x&=& -\frac{3}{p_\infty}(4p_\infty^4-1)\sqrt{1+p_\infty^2} b_0 \tau_1\,, \nonumber\\
B_1^y&=&  -2p_\infty \sqrt{1+p_\infty^2}(2p_\infty^2-1)(b^2-2p_\infty^2\tau_1^2)\,, \nonumber\\
B_2^y&=& -\frac{(1+p_\infty^2)(2p_\infty^2-1)^2}{p_\infty^2}(b^2-2p_\infty^2\tau_1^2)\,.
\eea
The no-log-part instead is given by
\bea
\tilde F^x_{\rm no-log}(\tau_1)&=& b_0\tau_1 \left[ \frac{C^x_{-5}}{{\mathcal D}^{10}}+\frac{C^x_{-4}}{{\mathcal D}^{8}}+\frac{C^x_{-3}}{{\mathcal D}^{6}} +\frac{C^x_{-2}}{{\mathcal D}^{4}}
\right.\nonumber\\
&+&\left.\frac{C^x_{-1}}{{\mathcal D}^{2}}+C^x_{0}
\right]\,,\nonumber\\
\tilde F^y_{\rm no-log}(\tau_1)&=& \sqrt{1+p_\infty^2} \left[ \frac{C^y_{-5}}{{\mathcal D}^{10}}+\frac{C^y_{-4}}{{\mathcal D}^{8}}+\frac{C^y_{-3}}{{\mathcal D}^{6}} +\frac{C^y_{-2}}{{\mathcal D}^{4}}\right.\nonumber\\
&+&\left. \frac{C^y_{-1}}{{\mathcal D}^{2}}+C^y_{0}+C^y_{1}{\mathcal D}^{2}
\right]\,,
\eea
where the  
coefficients $C_n^x$ and $C_n^y$ are listed in Table \ref{Table_coeff}.  
\begin{table}  
\caption{\label{Table_coeff} List of the   
coefficients entering the explicit expressions of the $x$ and $y$ components of the radiation-reaction force.}
\begin{ruledtabular}
\begin{tabular}{l|l}
$C^x_{-5} $ & $32p_\infty^{12}b_0^{10}(1+p_\infty^2)^2$\\
$C^x_{-4} $ & $-24 p_\infty^{10}b_0^8(5p_\infty^2+4)(1+p_\infty^2)$\\
$C^x_{-3} $ & $2p_\infty^8 b_0^6(100p_\infty^4+152p_\infty^2+53)$\\
$C^x_{-2} $ & $-p_\infty^6 b_0^4(184p_\infty^4+226p_\infty^2+53)$\\
$C^x_{-1} $ & $2p_\infty^4 b_0^2 (4p_\infty^4+32p_\infty^2 +3)$\\
$C^x_{0} $ & $-(4p_\infty^2+3)(2p_\infty^2-1)(1+p_\infty^2)$\\
\hline
$C^y_{-5} $ & $-32 b_0^{12}p_\infty^{11}(1+p_\infty^2)^2$\\
$C^y_{-4} $ & $8b_0^{10}p_\infty^9 (1+p_\infty^2)(17p_\infty^2+12)$\\
$C^y_{-3} $ & $-\frac23 b_0^8 p_\infty^7 (296p_\infty^4+440p_\infty^2+159)$\\
$C^y_{-2} $ & $\frac13 b_0^6 p_\infty^5 (292p_\infty^4+448p_\infty^2+159)$\\
$C^y_{-1} $ & $2b_0^4p_\infty^3 (2p_\infty^4-19p_\infty^2-3)$\\
$C^y_{0} $ & $-\frac{(2p_\infty^2-1)b_0^2 (22p_\infty^4+9p_\infty^2-9)}{3p_\infty}$\\
$C^y_{1} $ & $-\frac{2(2p_\infty^2-1)(3-5p_\infty^2)}{3p_\infty}$\\
\end{tabular}
\end{ruledtabular}
\end{table}

As a partial check on the coefficients listed in Table \ref{Table_coeff} we compared the 2PM radiation-reaction force ${\mathcal F}_{1\,\rm  rr}^\mu(\tau_1)$ 
with known results on the PN-expanded radiation-reaction force (in harmonic coordinates), and notably the 3.5PN accurate results of Refs.  \cite{Nissanke:2004er,Blanchet:2018yqa}. Working in the rest frame of particle 1 (at  the lowest-order approximation  where both particles move  on  straight lines) 
the check was done by inserting in the 3.5PN radiation-reaction acceleration, $A^{\le \rm 3.5PN}_{\rm rr}({\mathbf z}_1(t)-{\mathbf z}_2(t), {\mathbf v}_1(t),{\mathbf v}_2(t))$, the explicit  motions of both particles:
\bea
\label{v_def}
&&{\mathbf z}_1(t)-{\mathbf z}_2(t)=b_0{\mathbf e}_x -{\mathbf v}_2 t\,,\nonumber\\
&& {\mathbf v}_1=0\,,\qquad {\mathbf v}_2=- \frac{\sqrt{\gamma^2-1}}{\gamma} \,{\mathbf e}_y\equiv -v \,{\mathbf e}_y\,.
\eea
This (successful) check involved  the $O(G^2)$ part of $A_{\rm rr}^{\le \rm 3.5PN}$. Our 2PM accurate result for ${\mathcal F}_{a\,\rm  rr}^\mu(\tau_1)$ provides benchmarks fro checking future higher-PN-order computations of the radiation-reaction force in harmonic coordinates.

\begin{figure}[ht]
\includegraphics[scale=0.45]{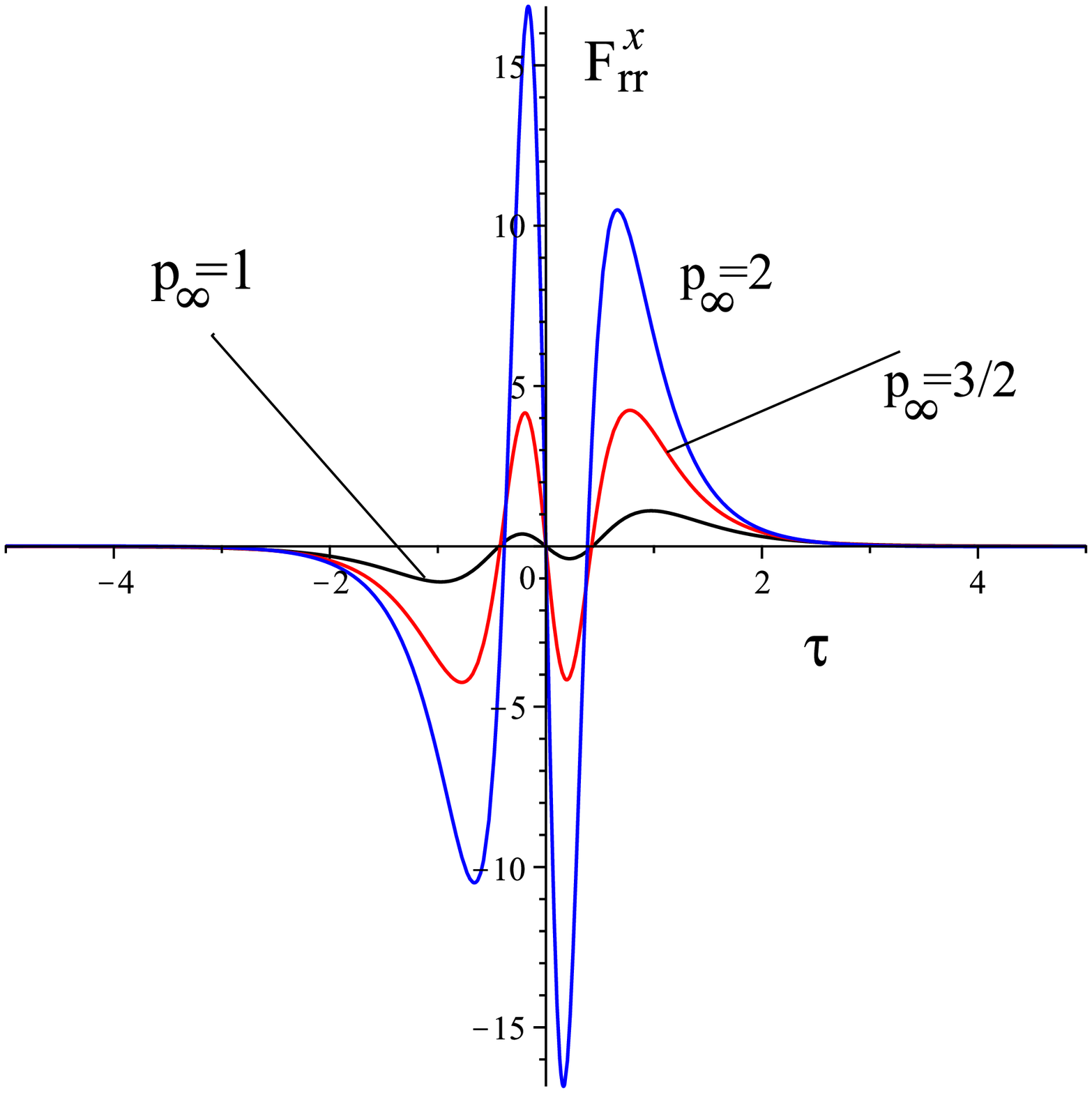}
\includegraphics[scale=0.45]{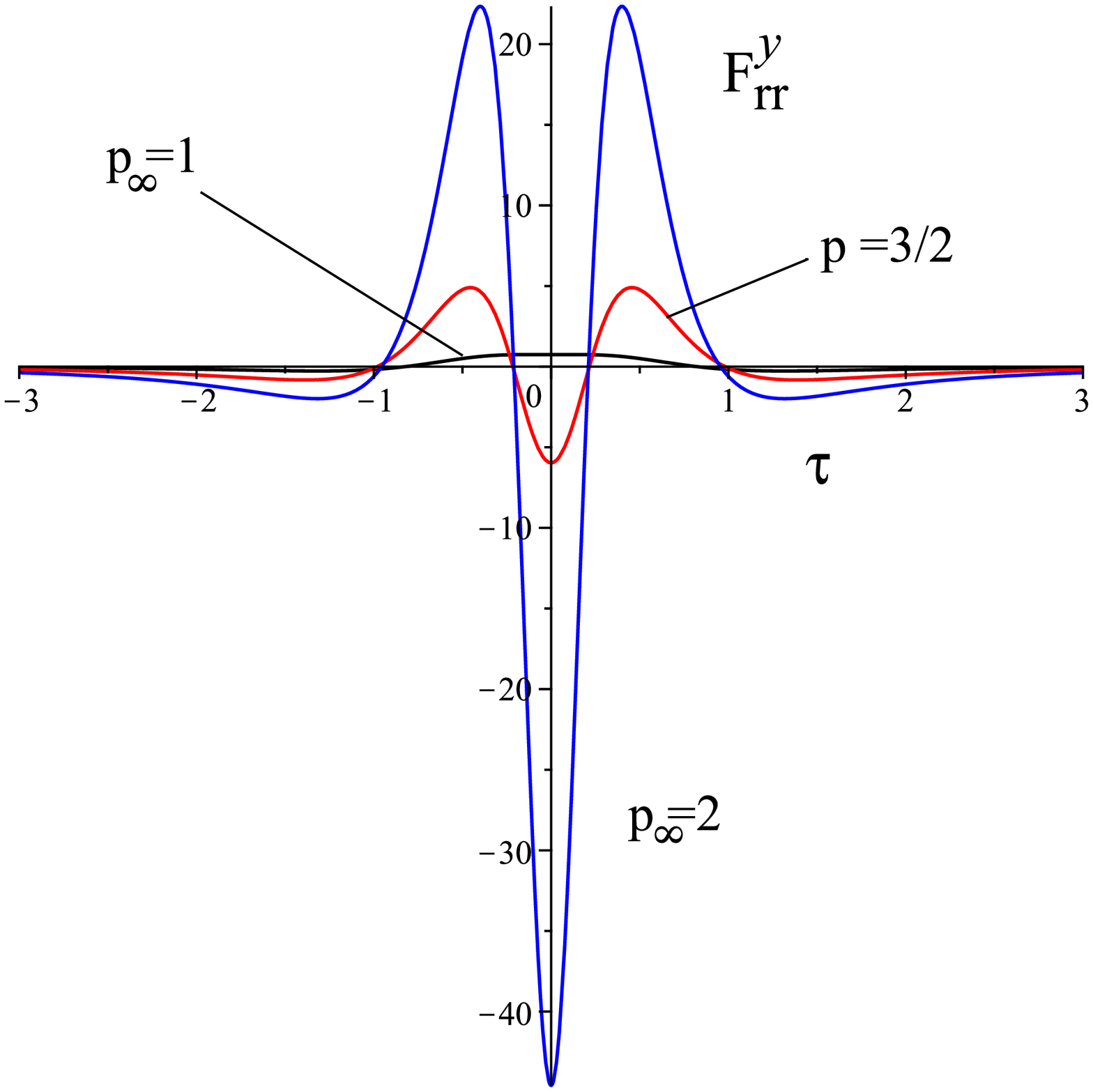}
\caption{\label{fig:F_rr_x} The $x$ and $y$ components of the radiation-reaction force acting on particle 1
(rescaled by $G^2 m_1^2 m_2$, and for $b_0=1$),
in its incoming rest-frame,  plotted as   functions of $\tau \equiv \tau_1$ for different values of $p_\infty$=[1 (black online), 3/2 (red online), 2 (blue online].   ${\mathcal F}_{1\,\rm  rr}^x$ is $\tau$-odd  whereas ${\mathcal F}_{1\,\rm  rr}^y$ is $\tau$-even. The asymptotic behaviors of the $x$ and $y$ components of the radiation-reaction forces are ${\mathcal F}_{1\, \rm rr}^x\sim \frac{\tau}{\tau^5}$ and ${\mathcal F}_{1\, \rm rr}^y\sim \frac{1}{|\tau|^3}$.}
\end{figure}

The two force components have simple properties under time-reversal, $\tau_1\to-\tau_1$. Namely
${\mathcal F}_{1\,\rm  rr}^x(\tau_1)$ is time-odd   while ${\mathcal F}_{1\,\rm  rr}^y(\tau_1)$ is time-even.
The two functions ${\mathcal F}_{1\,\rm  rr}^x(\tau_1)$ and ${\mathcal F}_{1\,\rm  rr}^y(\tau_1)$ are displayed in Fig. \ref{fig:F_rr_x}.

The midpoint (at $\tau_1=0$) values of the force components are
\bea
{\mathcal F}_{1\,\rm  rr}^x (0)&=&0\,,\nonumber\\
{\mathcal F}_{1\,\rm  rr}^y (0)&=&\frac{G^2m_1^2m_2}{b_0^3}\left[ \frac{(p_\infty^2+1) (2p_\infty^2-1)^2}{p_\infty^2} {\mathcal A}(p_\infty)\right.\nonumber\\
&-& 2 p_\infty(2 p_\infty^2-1) \sqrt{p_\infty^2+1}\ln(p_\infty^2+1)  \nonumber\\
&-&\left.
\frac{12 p_\infty^{10}-21 p_\infty^6-8 p_\infty^4-2 p_\infty^2+3}{3p_\infty (p_\infty^2+1)^{5/2}}  \right]\,.
\eea
In the limit  $\tau_1 \to \pm \infty$ we find that ${\mathcal F}_{1\,\rm  rr}^x$  vanishes as $\tau_1/|\tau_1|^5$ while 
${\mathcal F}_{1\,\rm  rr}^y $  vanishes as $1/|\tau_1|^3$ with   coefficients depending on $p_\infty$, namely
\bea
{\mathcal F}_{1\,\rm  rr}^x(\tau_1\to \pm \infty) &\approx &-\frac{G^2m_1^2m_2 b_0\tau_1}{|\tau_1|^5 p_\infty^3 } \left[12 (2 p_\infty^2+1)  {\mathcal A}(p_\infty)\right. \nonumber\\
&&\left.+ \frac{(4 p_\infty^2+3) (2 p_\infty^2-1) (p_\infty^2+1)}{p_\infty^2} \right]\nonumber\\
{\mathcal F}_{1\,\rm  rr}^y(\tau_1\to \pm \infty) &\approx &\frac{2G^2 m_1^2m_2(2p_\infty^2-1)\sqrt{1+p_\infty^2}}{p_\infty^2|\tau_1|^3}\times \nonumber\\
&&\left[ 4   {\mathcal A}(p_\infty)+\frac{(5 p_\infty^2-3)}{3p_\infty^2}   \right]\,.
\eea

As shown in Fig. \ref{fig:F_rr_x} the proper time evolution of the force components is rather complex, and involves (at least) two different time scales which behave differently in the $\gamma\to \infty$ limit. More precisely, while $D^2(\tau)=b_0^2+p_\infty^2 \tau^2$ involves the time scale $\tau_p=\frac{b_0}{p_\infty}$, the factor ${\mathcal D}^2(\tau)/\gamma^2=b_0^2+v^2 \tau^2$, where $v$ was defined in Eq. \eqref{v_def},
involves the time scale $\tau_v=\frac{b_0}{v}$. In the low-velocity limit $p_\infty\to 0$ the two time scales coincide and measure the usual characteristic Newtonian encounter time. In the high-energy limit $p_\infty \to \infty$ we have  $\tau_p\ll \tau_v$, due to relativistic blue-shift effects in retarded interactions.
In both limits we can write the two components of the force in terms of the rescaled time variable
\beq
\hat \tau =\frac{\tau}{\tau_p}\,.
\eeq
In the low-velocity limit ($\pinf \to 0$, $v \to 0$) we have
\bea
\frac{{\mathcal F}_{1\,\rm  rr}^x}{G^2m_1^2 m_2} &=&\frac{12}{5}p_\infty^3 \frac{\hat \tau}{(1+\hat \tau^2)^{5/2}}\,,\nonumber\\
\frac{{\mathcal F}_{1\,\rm  rr}^y}{G^2m_1^2 m_2} &=& \frac{4}{5}p_\infty^3 \frac{2\hat \tau^2-1}{(1+\hat \tau^2)^{5/2}}\,,
\eea
while, in the high-energy limit ($\pinf \to +\infty$, $v \to 1$), we have
\bea
\frac{{\mathcal F}_{1\,\rm  rr}^x}{G^2m_1^2 m_2} &=&-12 p_\infty^3 \ln(p_\infty) \frac{\hat \tau}{(1+\hat \tau^2)^{5/2}}\,,\nonumber\\
\frac{{\mathcal F}_{1\,\rm  rr}^y}{G^2m_1^2 m_2} &=& 4 p_\infty^4 \ln(p_\infty) \frac{2\hat \tau^2-1}{(1+\hat \tau^2)^{5/2}}\,. 
\eea
Note that, apart from different prefactors (including varying signs), each component involves the same function of $\hat \tau$ in both limits.

\subsection{Moments of the $O(G^2)$ radiation-reaction force}

The integrated value  of the force vanishes, 
\beq
\label{zero_mom}
\int d\tau_1 {\mathcal F}^{\mu}_{1\rm rr}(\tau_1) =0\,,
\eeq
as expected from the conservation of total linear momentum up to the $G^3$ level. See also further discussion below.

Let us consider  the integrated moments of the radiation-reaction force, i.e., the integrals
\beq
\label{uno_mom}
I_{a}^{(n)\alpha}\equiv \int_{-\infty}^\infty d\tau \, \tau^n \, {\mathcal F}_{a\rm rr}^\alpha(\tau)\,.
\eeq
For $n=0$ we have $I_{1}^{(0)\alpha}=0+O(G^3)$, 
in view of Eq. \eqref{zero_mom}. 
As we  shall see below of particular importance is the first moment  $n=1$ which is found to be
\bea
\label{I1_1}
I_{1}^{(1)t}&=&0\,,\nonumber\\
I_{1}^{(1)x}&=&\frac{G^2 m_1^2 m_2}{b_0}\, c_I(v)\,  {\mathcal I}(v)\,,
\nonumber\\
I_{1}^{(1)y}&=&0\,.
\eea
Here, we defined  (remembering the definition of $v$, Eq. \eqref{v_def}),
\beq
\label{cI_def}
c_I(v)\equiv \frac{1+v^2}{v\sqrt{1-v^2}}=\frac{2\gamma^2-1}{\sqrt{\gamma^2-1}}\,,
\eeq
and, consistently with Eqs. (4.7) and (4.8) of Ref. \cite{Damour:2020tta},
\bea
\label{calI_def}
{\mathcal I}(v)&\equiv &-\frac{16}{3}+\frac{2}{v^2}+\frac{2(3v^2-1)}{v^3}{\mathcal A}(v)\,,\nonumber\\
{\mathcal A}(v)&\equiv &{\rm arctanh}v=\frac12 \ln \frac{1+v}{1-v}\nonumber\\
&=&2{\rm arcsinh}\left(\sqrt{\frac{\gamma-1}{2}}\right)\,.
\eea
Note that ${\mathcal A}(v)$ is numerically equal to the ${\mathcal A}(p_\infty)$  defined in Eq. \eqref{mathcal_A} above.

In terms of $p_\infty$  our results read 
\bea
I_{1}^{(1)x}&=&\frac{2G^2 m_1^2 m_2(2p_\infty^2+1)}{b p_\infty^3}
\left[
\frac{(2p_\infty^2-1)\sqrt{1+p_\infty^2}}{p_\infty}\times \right.\nonumber\\
&&\left.\times {\mathcal A}(p_\infty)-\frac{5p_\infty^2-3}{3}
\right]\,.
\eea
Recalling that $e_x^\mu=\hat b_{12}^\mu$ we can rewrite the above relation in a covariant form
\beq
I_{1}^{(1)\mu}=  \frac{G^2 m_1^2 m_2}{b_{12}^2}b_{12}^\mu  c_I(v)\, {\mathcal I}(v)\,.
\eeq

\subsection{Schott term in the radiation-reaction force}
We have seen above that ${\mathcal F}^{\mu}_{1 \rm rr}(\tau_1)$ starts at $O(G^2)$
but that
\beq
\label{int_F_eq_0}
\int d\tau_1 {\mathcal F}^{\mu}_{1\rm rr}(\tau_1) =O(G^3)\,.
\eeq
This situation is similar to the well-known structure of the Abraham-Lorentz-Dirac radiation reaction force for a test-charge particle in an external field, namely (with $ \cdot \equiv \frac{d}{d \tau}$)
\beq
\label{moto_em}
m_1 \frac{du^\mu}{d\tau}=\frac23 e_1^2  ( \ddot u^\mu - \dot u^2 u^\mu)+F^\mu_{\rm ext}\equiv {\mathcal F}^{\mu}_{\rm rr}+F^\mu_{\rm ext}\,.
\eeq 
Only the second term, $-\frac23 e_1^2 \dot u^2 u^\mu$, in $ {\mathcal F}^{\mu}_{\rm rr}$ is linked to the emission of radiation. Indeed, 
\beq
P^\mu_{\rm rad}(\tau)=+\frac23 e_1^2 \int_{-\infty}^\tau d\tau \dot u^2 u^\mu\,,
\eeq
or
\beq
\frac{d P^\mu_{\rm rad}}{d\tau}=+\frac23e_1^2 \dot u^2 u^\mu\,.
\eeq
The first term   $\frac23 e_1^2  \ddot u^\mu$ in $ {\mathcal F}^{\mu}_{\rm rr}$ is a total time derivative. It was interpreted by Schott~\cite{Schott_book}
as part of the interaction energy between the charge and the field.
Defining 
\beq
P^\mu_{\rm kin}=m_1 u^\mu\,,
\eeq
and
\beq
P^\mu_{\rm Schott}=-\frac23 e_1^2 \dot u^\mu
\eeq
Eq.  \eqref{moto_em} reads
\beq
\label{eq_P_tot}
\frac{dP^\mu_{\rm tot}}{d\tau}=F^\mu_{\rm ext}\,,
\eeq
where 
\beq
\label{eq_P_tot2}
P^\mu_{\rm tot}= P^\mu_{\rm kin}+P^\mu_{\rm Schott}+ P^\mu_{\rm rad}\,.
\eeq
Eqs. \eqref{eq_P_tot} and \eqref{eq_P_tot2} exhibit the role of the Schott momentum as a necessary additional contribution 
in the energy-momentum balance between external force,  particle and radiation. If the external force is due to the electromagnetic interaction between the test charge $e_1$ and another (heavy) charge $e_2$ (with $e_2\sim e_1$), the Schott momentum and the radiated momentum scale differently with the coupling constant 
$\alpha\equiv e_1 e_2 \sim e_1^2$. While $P^\mu_{\rm rad}=O(\alpha^3)$, the Schott momentum scales with a lower power of $\alpha$: $P^\mu_{\rm Schott}=O(\alpha^2)$.
Correspondingly, in the radiation-reaction force, the Schott term is $O(\alpha^2)$, while the \lq\lq proper" radiation-reaction term, $-\frac23 e_1^2 \dot u^2 u^\mu$, is proportional to $\alpha^3$. 

The latter situation is analogous to the structure of the radiation-reaction force in gravity, with the analogy $\alpha\leftrightarrow  G$.
The $O(G^2)$ radiation-reaction force discussed in the present paper is analogous to the Schott force $\frac23 e_1^2 \ddot u$.
It would be natural to decompose the gravitational radiation-reaction force as
\beq
{\mathcal F}^{\mu}_{\rm rr}=-\frac{dP^\mu_{\rm Schott}}{d\tau}+{\mathcal F}^{\mu}_{\rm rr\, proper}\,,
\eeq
with $P^\mu_{\rm Schott}=O(G^2)$ and ${\mathcal F}^{\mu}_{\rm rr\, proper}=O(G^3)$. In this relation  only be ${\mathcal F}^{\mu}_{\rm rr\, proper}$ would be responsible for the radiative loss of linear momentum of each particle.
Our present treatment is limited to $O(G^2)$ accuracy, and therefore only gives access to $P^\mu_{\rm Schott}$ to this order.
Integrating 
\beq
\label{eq_F_1_schott}
{\mathcal F}^\mu_{1\, \rm rr}(\tau_1)=-\frac{d}{d\tau_1} P^{\mu}_{1\,\rm Schott}+O(G^3)\,,
\eeq
we find that the two components of  $P^\mu_{1\,\rm Schott}(\tau_1)$,
\beq
P^\mu_{1\,\rm Schott}(\tau_1)=P^{x}_{1\,\rm Schott}(\tau_1)e_x^\mu+P^{y}_{1\,\rm Schott}(\tau_1)e_y^\mu\,,
\eeq
are given, at order $G^2$, by the following expressions
\begin{widetext}
\bea
\frac{ P_{1\,\rm Schott}^{x\, G^2}}{G^2 m_1^2m_2}&=&
-2 \frac{b_0}{D^3} (1+2 p_\infty^2)\ln\left(\frac{{\mathcal D}^2}{D^2}\right)
-\frac{3}{4 b_0^2 p_\infty} (4+5 p_\infty^2){\rm arctan}\left(\frac{p_\infty b_0}{D}\right)\nonumber\\
&+&\frac{b_0(4p_\infty^4-1)}{p_\infty^3 D^3} \left[\sqrt{1+p_\infty^2}{\rm arcsinh}(p_\infty)-p_\infty 
\right]\nonumber\\
&+&\frac{1}{Db_0p_\infty}\left[2(1+6p_\infty^2+4p_\infty^4)
-\frac14(28+84p_\infty^2+47p_\infty^4)\frac{D^2}{{\mathcal D}^2}\right.\nonumber\\
&+&\left.\frac16 (74+136 p_\infty^2+65p_\infty^4)\frac{D^4}{{\mathcal D}^4}
-\frac23 (1+p_\infty^2)(17+11p_\infty^2)\frac{D^6}{{\mathcal D}^6}
+4(1+p_\infty^2)^2\frac{D^8}{{\mathcal D}^8}
\right]\,,\nonumber\\
\frac{ P_{1\,\rm Schott}^{y\, G^2}}{G^2 m_1^2m_2\tau}&=&\frac{2p_\infty (2p_\infty^2-1) \sqrt{1+p_\infty^2}}{D^3}\ln\left(\frac{{\mathcal D}^2}{D^2}\right)
+\frac{\sqrt{1+p_\infty^2}(2p_\infty^2-1)^2}{p_\infty^2 D^3}[-\sqrt{1+p_\infty^2}{\rm arcsinh}(p_\infty)+p_\infty]
\nonumber\\
&+& \frac{\sqrt{1+p_\infty^2}}{D^3}\left[
-2p_\infty (4p_\infty^2-1)\frac{D^2}{{\mathcal D}^2}
+p_\infty(6p_\infty^2-5)\frac{ D^4 }{{\mathcal D}^4}
+\frac{2}{3}p_\infty (11+8p_\infty^2)\frac{ D^6}{{\mathcal D}^6}
-4p_\infty (1+p_\infty^2)\frac{ D^8}{{\mathcal D}^8}
\right]\,.\nonumber\\
\eea
\end{widetext}
Remembering the definitions of $D(\tau_1)$ (see the first of Eqs. \eqref{D_and_cal_D_defs}) and ${\mathcal D}(\tau_1)$ (see the second of Eqs.  \eqref{D_and_cal_D_defs}),
one sees that $ P_{\rm Schott}^{x\, G^2}$ is an even function of $\tau$ while $ P_{\rm Schott}^{y\, G^2}$ is an odd function of $\tau$. In addition, when $\tau \to \pm \infty$, both components of $P_{1\,\rm Schott}^{y\, G^2}$ tend to zero. More precisely:
\beq
 P_{1\,\rm Schott}^{x\, G^2} \sim \frac{1}{|\tau|^3}\,,\qquad  P_{1\,\rm Schott}^{y\, G^2} \sim \frac{\tau}{|\tau|^3}\,.
\eeq

In the following we will study the role of the radiation-reaction force ${\mathcal F}_{1\, \rm rr}^\mu$ on the conservation (respectively, dissipation) of the linear momentum  (respectively, angular momentum) of the system.

\section{Evolution of Noetherian quantities at $O(G^2)$}

As discussed in Section \ref{time_sym_dyn} the time-symmetric dynamics of two masses admits conserved Noetherian quantities associated with the Poincar\'e symmetry of its Fokker action.
As explicitly shown, at the 1PM level, in Eqs. \eqref{P_and_J}  
the Noetherian conserved quantities $P_{\rm sys}^\mu$ and $J_{\rm sys}^{\mu\nu}$ are the sum of kinematical quantities and interaction contributions. 
The existence of these conserved quantities for the time-symmetric dynamics, and the decomposition of the retarded force given in Eq. \eqref{deco_F}, show that when considering the retarded dynamics  the non-conservation of $P_{\rm sys}^\mu$ and $J_{\rm sys}^{\mu\nu}$ {\it will only come} from the additional term  ${\mathcal F}_{a\, \rm rr}^\mu$ in the equations of motion. A way to make this explicit would be (following Lagrange's method of variation of constants)  to express $P_{\rm sys}^\mu$ and $J_{\rm sys}^{\mu\nu}$ as functions of four quantities 
$z_1^\mu $, $u_1^\mu$, $z_2^\mu$ and $u_2^\mu$, which can serve as \lq\lq initial" conditions determining a solution of the time-symmetric equations of motion.
Then, starting from the functions $P_{\rm sys}^\mu(z_1,u_1,z_2,u_2)$ and $J_{\rm sys}^{\mu\nu}(z_1,u_1,z_2,u_2)$ we get evolution equations for these quantities under the retarded dynamics of the form
 \bea
\label{td_sys}
dP_\mu^{\rm sys}&=&\sum_a  \frac{\partial P_\mu^{\rm sys}}{\partial u_a^\lambda} \frac{{\mathcal F}^\lambda_{a\, \rm rr}(\tau_a)}{m_a}\frac{d\tau_a}{d\sigma} d\sigma\,, \nonumber\\
dJ_{\mu\nu}^{\rm sys}&=&\sum_a  \frac{\partial J_{\mu\nu}^{\rm sys}}{\partial u_a^\lambda} \frac{{\mathcal F}^\lambda_{a\, \rm rr}(\tau_a)}{m_a}\frac{d\tau_a}{d\sigma} d\sigma \,.
\eea
Here, $\sigma$  parametrizes a way to correlate the common sliding of the data $z_a,u_a$ along the two world lines.
For instance, one could use (as is done when dealing with the PN-expanded dynamics) a coordinate time $t$ is some Lorentz frame.

When working in a PM-expanded way, the facts that the Noetherian quantities differ from kinematical quantities by $O(G)$ interaction contributions    
 and that the radiation-reaction force starts at order $O(G^2)$ allows us to write the following $G^2$-accurate evolution equations\footnote{When working at order $G^3$ one would need to take into account the $u_a^\mu$-dependence of the interaction contributions to the Noetherian quantities.}
 \bea
dP_\mu^{\rm sys}&=&\sum_a  \frac{\partial P_\mu^{\rm kin}}{\partial u_a^\lambda} \frac{{\mathcal F}^\lambda_{a\, \rm rr}(\tau_a)}{m_a}\frac{d\tau_a}{d\sigma} d\sigma +O(G^3)\,,\nonumber\\
dJ_{\mu\nu}^{\rm sys}&=&\sum_a  \frac{\partial J_{\mu\nu}^{\rm kin}}{\partial u_a^\lambda} \frac{{\mathcal F}^\lambda_{a\, \rm rr}(\tau_a)}{m_a}\frac{d\tau_a}{d\sigma} d\sigma +O(G^3) \,,
\eea
i.e., explicitly
 \bea
\label{dP_dJ_expl}
dP^\mu_{\rm sys}&=&\sum_a  {\mathcal F}^\mu_{a \, \rm rr}(\tau_a) \frac{d\tau_a}{d\sigma} d\sigma +O(G^3)\,,\nonumber\\
dJ^{ \mu\nu}_{\rm sys} &=&\sum_a (z_{a}^\mu  {\mathcal F}^\nu_{a\, \rm rr}(\tau_a)-z_{a}^{\nu} {\mathcal F}^\mu_{a\, \rm rr}(\tau_a) )\frac{d\tau_a}{d\sigma} d\sigma\nonumber\\
& +&O(G^3) \,.
\eea

\subsection{Linear momentum}

In Appendix \ref{lin_mom_A} we evaluate at 1PM accuracy the interaction contribution to the total linear momentum of a binary system.  Our computation  
explicitly shows that the interaction contribution to $P_\mu^{\rm sys}$ vanishes both in the asymptotic incoming and outgoing states. In addition, 
it also gives an explicit check of the  
conservation of $P_\mu^{\rm sys}$ under the 1PM dynamics (which is time-symmetric by itself at this order, as discussed above).

When working at the 2PM accuracy with the retarded dynamics, the first equation in Eqs. \eqref{dP_dJ_expl} says that the total change during scattering of the Noetherian conserved momentum of the system is given by
\beq
[P_{\rm sys}^\mu]_{-\infty}^{+\infty}=\sum_a \int d\tau_a {\mathcal F}^\mu_{a \, \rm rr}(\tau_a) +O(G^3)\,,
\eeq
where we adopted the notation $[Q]_{-\infty}^{+\infty}\equiv Q(+\infty)--Q(-\infty)$.

Using Eq. \eqref{int_F_eq_0} this yields
\beq
\label{Delta_P_sys}
[P^{\rm sys}]_{-\infty}^{+\infty}=O(G^3)\,.
\eeq
In addition, using the fact that the interaction contribution to $P^{\rm sys}$ vanishes in the asymptotic states, Eq. \eqref{Delta_P_sys},  we get  
\beq
[m_1 u_1+m_2u_2]_{-\infty}^{+\infty}=O(G^3)\,.
\eeq

\subsection{Angular momentum}

In Appendix \ref{ang_mom_A} we  evaluate at 1PM accuracy the interaction contribution to the total angular momentum of a system to 
explicitly check its conservation under the 1PM dynamics (which is time-symmetric by itself at this order, as discussed above).
Contrary to the case of the linear momentum, the interaction contribution to the angular momentum  {\it does not vanish} in the asymptotic limits $t\to \pm \infty$.
However, it exactly compensates opposite contributions (linked to the asymptotic logarithmic behaviors of the world lines) present in the kinematical part of the angular momentum.
 
The non-conserved, and asymptotically non-vanishing, contributions to $J_{\mu\nu}^{\rm kin}$ were called \lq\lq scoot terms" 
in Refs. \cite{Gralla:2021eoi,Gralla:2021qaf}. The scoot contributions to $J_{\mu\nu}^{\rm kin}$ are not Lorentz-invariant (see Eqs. \eqref{small_j_mu_nu}, \eqref{j_int_plus} and \eqref{j_int_minus}
 in Appendix \ref{ang_mom_A}). Our computations in Appendix \ref{ang_mom_A} explicitly show that these non Lorentz-invariant and non-conserved scoot kinematical contributions cancel against corresponding  (Fokker) interaction contributions to the total angular momentum.

The final (manifestly Poincar\'e-invariant, and conserved) result for $J_{\mu\nu}^{\rm sys}(\tau_1,\tau_2)=J_{\mu\nu}^{\rm kin}(\tau_1,\tau_2)+J_{\mu\nu}^{\rm int}(\tau_1,\tau_2)$ is
\bea
\label{J_sys_final}
J_{\mu\nu}^{\rm sys}(\tau_1,\tau_2)
&=& \left( (z_1(0)+\delta^G z_1)\wedge p_1^-\right)_{\mu\nu}\nonumber\\
&+&\left( (z_2(0)+\delta^G z_2)\wedge p_2^-\right)_{\mu\nu}\nonumber\\
&+&O(G^2)\,,
\eea
where
\bea
\delta^G z_1^\mu&=& +Gm_2 \frac{2\gamma^2-1}{(\gamma^2-1)}\hat b_{12}^\mu\,,\nonumber\\
\delta^G z_2^\mu&=& -Gm_1 \frac{2\gamma^2-1}{(\gamma^2-1)}\hat b_{12}^\mu\,.
\eea
In other words, defining $b^\mu_{a \, \rm in} \equiv z_a^\mu(0)+\delta^G z_a^\mu$,
we can write $J_{\mu\nu}^{\rm sys}(\tau_1,\tau_2)$ as
\bea
\label{J_sys_final2}
J_{\mu\nu}^{\rm sys}(\tau_1,\tau_2)
&=& (b_{1 \,\rm in}\wedge p_1^-)_{\mu\nu}\nonumber\\
&+&(b_{2 \, \rm in}\wedge p_2^-)_{\mu\nu}\nonumber\\
&+&O(G^2)\,,
\eea
The modulus of the difference $b^\mu_{1 \, \rm in} - b^\mu_{2 \, \rm in}$ is 
equal to the incoming impact parameter $b_{\rm in}$ defined in Eq. \eqref{impact_eq}.

When working at the 2PM accuracy with the retarded dynamics, the second equation in Eqs. \eqref{dP_dJ_expl}  says that the total change during scattering of the Noetherian conserved angular momentum of the system is given by
\beq
\label{J_rad_eq_fin}
[J_{\rm sys}^{\mu\nu}]_{-\infty}^{+\infty}=\sum_a \int d\tau_a (z_a(\tau_a)\wedge {\mathcal F}_{a \, \rm rr}(\tau_a))^{\mu\nu} +O(G^3)\,.
\eeq
At order $G^2$ it is enough to insert the straight line approximation for $z_a^\mu(\tau_a)$ in Eq. \eqref{J_rad_eq_fin},
\beq
\label{J_rad_eq_fin2}
[J_{\rm sys}^{\mu\nu}]_{-\infty}^{+\infty}=\sum_a \int d\tau_a [(z_a(0)+\bar u_a\tau_a)\wedge {\mathcal F}_{a \, \rm rr}(\tau_a)]^{\mu\nu} +O(G^3)\,.
\eeq
The contribution proportional to $z_a(0)$ vanishes because of Eq. \eqref{zero_mom}. 
By contrast, the contribution proportional to $\bar u_a \tau_a$ yields a term proportional to the first moment of the radiation-reaction force: 
\beq
I_{a}^{(1)\mu}=\int d\tau_a \tau_a {\mathcal F}_{a \, \rm rr}^\mu(\tau_a)\,,
\eeq
leading to
\beq
\label{J_rad_eq_fin3}
[J_{\rm sys}^{\mu\nu}]_{-\infty}^{+\infty}=\sum_a [ \bar u_a \wedge I_{a}^{(1)}]^{\mu\nu} +O(G^3)\,.
\eeq

We have evaluated the moment $I_{a}^{(1)\mu}$  in Eq. \eqref{uno_mom} above.
This yields the central result of the present paper~\footnote{The decomposition of $[J_{\rm sys}^{\mu\nu}]_{-\infty}^{+\infty}$ in two terms associated with the two particles arises out technically at our present $G^2$ order, but is not expected to be natural at higher PM orders.}:
\beq
\label{J_rad_eq_fin4}
[J_{\rm sys}^{\mu\nu}]_{-\infty}^{+\infty}=\Delta J_1^{\mu\nu}+\Delta J_2^{\mu\nu}
 +O(G^3)\,,
\eeq
where
\bea
\Delta J_1^{\mu\nu}&=& \bar u_1^\mu  \wedge \int d\tau_1 \tau_1 {\mathcal F}_{1 \, \rm rr}^\nu(\tau_1) \nonumber\\
&=&\frac{G^2 m_1 m_2}{b_{12}^2}\, c_I(v)\, {\mathcal I}(v)\, [p_1\wedge b_{12}]^{\mu\nu}\,,
\eea
and  
\bea
\Delta J_2^{\mu\nu}&=& \bar u_2^\mu \wedge \int d\tau_2 \tau_2 {\mathcal F}_{1 \, \rm rr}^\nu(\tau_2)\nonumber\\
&=&\frac{G^2 m_1 m_2}{b_{12}^2}\, c_I(v)\, {\mathcal I}(v)\, [p_2\wedge b_{21}]^{\mu\nu}\,,
\eea
leading to
\beq
\label{J_rad_eq_fin5}
[J_{\rm sys}^{\mu\nu}]_{-\infty}^{+\infty}=\frac{G^2 m_1 m_2}{b_{12}^2}\, c_I(v)\, {\mathcal I}(v)\, [(p_1-p_2)\wedge b_{12}]^{\mu\nu} 
 +O(G^3)\,.
\eeq
This total variation in the Noetherian angular momentum of the binary system coincides with the opposite
of the  $O(G^2)$ integrated radiative flux  of angular momentum computed in Refs. \cite{Damour:2020tta,Manohar:2022dea}.  It was obtained here by a direct equations-of-motion-based approach (similar to the lowest PN order of Ref. \cite{Damour:1981bh}) without ever appealing to a balance with fluxes of angular momentum in the radiation zone. See Concluding Remarks for further discussion.

\subsection{Radiation-reaction-induced shifts in the world lines of the two bodies}

In view of the conservative-plus-dissipative  decomposition of the equations of motion
\beq
m_a \frac{d^2}{d\tau_a^2}z_{a}^\mu(\tau_a)={\mathcal F}_{aR}^\mu(\tau_a)={\mathcal F}_{aS}^\mu(\tau_a)+{\mathcal F}_{a\,\rm rr}^\mu(\tau_a)\,,
\eeq
one can, at order $G^2$, accordingly decompose the solution world lines for the retarded dynamics in conservative and radiation-reaction parts:
\beq
z_{a}^\mu(\tau_a)=z_{a S}^{\mu}(\tau_a)+z_{a\,\rm rr}^\mu(\tau_a)+O(G^3)\,.
\eeq
Here, $z_{a S}^{\mu}(\tau_a)$ is the solution of the conservative dynamics while the radiation-reaction shift $z_{a\,\rm rr}^\mu(\tau_a)$
is the solution of 
\beq
\label{eom_rr}
m_a \frac{d^2}{d\tau_a^2}z_{a\, \rm rr}^\mu(\tau_1)={\mathcal F}_{a\, \rm rr}^\mu(\tau_1)\,.
\eeq
Here and below we suppress the $O(G^3)$ error terms.
Taking into account Eq. \eqref{eq_F_1_schott} we can integrate Eq. \eqref{eom_rr} once obtaining
\beq
\label{dz_sch}
m_a \frac{d}{d\tau_a}  z_{a\, \rm rr}^\mu(\tau_a)=-P_{a\,\rm  Schott}^\mu(\tau_a)\,,
\eeq
where we imposed the boundary condition  that $\frac{d z_{a\, \rm rr}^\mu(\tau_a)}{d\tau_a}$ vanishes in the incoming state.
As we have shown above that $P_{a\,\rm  Schott}^\mu(\tau_a)$ vanishes in both asymptotic
limits, $\tau_a\to \pm \infty$, we see that $\frac{d z_{a\, \rm rr}(\tau_a)}{d\tau_a}^\mu$ vanishes also in the outgoing state.

Integrating now Eq. \eqref{dz_sch} we find that the radiation-reaction-induced shift  of each world line is equal to
\beq
z_{a\, \rm rr}^\mu(\tau_a)=-\frac{1}{m_a}\int_{-\infty}^{\tau_a} d\bar \tau_a P_{a\,\rm  Schott}^\mu(\bar \tau_a)\,,
\eeq 
where we imposed the condition that $z_{a\, \rm rr}^\mu(\tau_a)$ vanishes in the incoming state.

Taking the limit $\tau_a\to +\infty$ and using Eqs. \eqref{I1_1} we find that the radiation-reaction shifts in the {\it outgoing} world lines  
are given by
\bea
\label{delta_rr_za}
[z_{1\, \rm rr}^\mu]^{+\infty}&=& -\alpha\, \hat b_{12}^\mu\,,\nonumber\\
{}[z_{2\, \rm rr}^\mu]^{+\infty}&=& -\alpha\, \hat b_{21}^\mu= + \alpha\, \hat b_{12}^\mu\,,
\eea
where 
\beq
\alpha\equiv \frac{G^2m_1m_2}{b_{12}}c_I(v)\, {\mathcal I}(v)\,.
\eeq
Note that $\alpha$ is positive so that each world line is shifted towards  the other world line.

These shifts can be interpreted in terms of an outgoing impact parameter $b^{\rm out}$ that differs (because
of radiation-reaction effects) from the incoming one $b^{\rm in}$ (such that $J_{\rm c.m.}^{\rm in}=b^{\rm in} P_{\rm c.m.}^{\rm in}$, see Eq.\eqref{Jcm}), by
\beq
b^{\rm out}=b^{\rm in}-2\alpha\,.
\eeq 
Such a shift implies a c.m. angular momentum decrease    
\beq
J_{\rm c.m.}^{\rm out}=J_{\rm c.m.}^{\rm in}-2\alpha(v) P_{\rm c.m.}\,,
\eeq 
which agrees with the result of Ref. \cite{Damour:2020tta}, and is easily seen to be compatible with  
the Poincar\'e-covariant result, Eq. \eqref{J_rad_eq_fin5}.
We leave to future investigations a discussion of the relation of the individual radiation-reaction 
worldline shifts, Eqs. \eqref{delta_rr_za},  to the recent corresponding results of Ref. \cite{DiVecchia:2022piu}.

\section{Concluding remarks}

We computed the effect of radiation-reaction at the second post-Minkowskian order $O(G^2)$.
The radiation-reaction force was defined by comparing the 2PM-accurate retarded dynamics, Eqs. \eqref{master_eq} and \eqref{master_eq2},  to its time-symmetric counterparts, Eqs. \eqref{FSvsRA} and \eqref{calFSvsRA}.
This led to the definition \eqref{def_f_rr_fin} of the radiation-reaction force ${\mathcal F}_{a\, \rm rr}^\mu(\tau_a)$. The explicit value of  ${\mathcal F}_{1\,\rm rr}^\mu(\tau_1)$ is given in subsection \ref{exp_expr_g2rr}.

Capitalizing on the existence of Noetherian conserved quantities, under the Fokker-Wheeler-Feynman-type time-symmetric dynamics  for the binary system, $P_{\rm sys}^\mu(z_1,u_1,z_2,u_2)$ and $J_{\rm sys}^{\mu\nu}(z_1,u_1,z_2,u_2)$,  we used the method of variation of constants to compute 
the evolution of $P_{\rm sys}^\mu(z_1,u_1,z_2,u_2)$ and $J_{\rm sys}^{\mu\nu}(z_1,u_1,z_2,u_2)$ under the retarded dynamics, see Eqs. \eqref{td_sys}
and \eqref{dP_dJ_expl}. 

Consistently with current knowledge we found that the total linear momentum of the system is conserved at the 2PM order.
By contrast, we found that the total variation of the angular momentum of the system under the retarded dynamics was given by Eq. \eqref{J_rad_eq_fin5}, namely
\bea
\label{J_rad_eq_fin6}
J_{\rm sys, out}^{\mu\nu}-J_{\rm sys, in}^{\mu\nu} &=&\frac{G^2 m_1 m_2}{b_{12}^2}\, c_I(v) {\mathcal I}(v) [(p_1-p_2)\wedge b_{12}]^{\mu\nu}\nonumber\\ 
& +&O(G^3)\,,
\eea
where $ c_I(v)$ and ${\mathcal I}(v)$ are respectively defined in Eqs. \eqref{cI_def} and \eqref{calI_def}.

The crucial point in our derivation of the result \eqref{J_rad_eq_fin6} is that it was obtained here directly from the (near-zone, retarded) mechanical equations of motion of the two world lines,   without ever evaluating fluxes of angular momentum in the radiation zone. 
This allows us to by-pass the subtleties linked to the definition of angular momentum at future null infinity with its attendant  Bondi-Metzner-Sachs-related super-translation  ambiguities \cite{Veneziano:2022zwh,Porrati2022}.
 
Our derivation is a generalization to all powers of $v/c$ of the lowest order result of Ref. \cite{Damour:1981bh} which was also directly based on the $G^2$-accurate retarded equations of motion of the  binary system.

We expect that our direct  equations-of-motion-based approach can be extended to the $G^3$ level, where the Fokker action should be well defined.
By contrast, several arguments (presence of tails \cite{Bini:2021gat,Dlapa:2022lmu,Bini:2022enm}, effects proportional the square of ${\mathcal F}_{a \, \rm rr}^\mu$) suggest that the $G^4$ level will introduce new subtleties.

\section*{Acknowledgments}
T.D. thanks Rodolfo Russo and Gabriele Veneziano for informative discussions. 
The present research was partially supported by the  ``2021 Balzan Prize for Gravitation: Physical and
Astrophysical Aspects", awarded to Thibault Damour.
D.B. thanks ICRANet for partial support, and acknowledges sponsorship of the Italian Gruppo Nazionale per la Fisica Matematica (GNFM) of the Istituto Nazionale di Alta Matematica (INDAM).
D.B. also acknowledges the highly stimulating environment of the Institut des Hautes Etudes Scientifiques.

\appendix

\section{Definition of retarded/advanced quantities}
\label{App:retarded}

Advanced and retarded definitions can be treated together with an indicator $\epsilon$, where $\epsilon=1$ in the retarded case and $\epsilon=-1$ in the advanced one. Using the notation of Ref. \cite{Bel:1981be}, which deals  both with a generic field point $x$  and several   world line points ($z_1$, $\hat z_{2\epsilon}$, etc.), we have~\footnote{Note that $u\cdot (x-z_\epsilon)=- u^0(t-z^0_\epsilon) + {\bf u} \cdot ({\bf x}- {\bf z_\epsilon})$ is negative in the retarded case 
($\epsilon=+1$), but positive in the advanced case ($\epsilon=-1$), while $w_\epsilon$ and $\omega_\epsilon$   
are negative in both cases.}  
\begin{eqnarray}
w_\epsilon &=& u_1 \cdot u_{2\epsilon}\,,\qquad \qquad \qquad 
\omega_\epsilon =u_1  \cdot \hat u_{2\epsilon}\nonumber\\
r_\epsilon &=& -\epsilon u_1 \cdot (x-z_1 ) \,,\qquad\,\,\,\, \rho_\epsilon = -\epsilon (z_1-\hat z_{1\epsilon})\cdot \hat u_{2\epsilon} \nonumber\\
n_\epsilon &=&   
\frac{1}{r_\epsilon}P(u_1)(x-z_1)\,,\qquad 
\nu_\epsilon =  
\frac{1}{\rho_\epsilon}P(\hat u_{2\epsilon})(z_1-\hat z_{2\epsilon})\,,\nonumber\\
\end{eqnarray}
to which one has to add
\bea
A_\epsilon^\alpha &=&\frac{1}{\rho_\epsilon}[P(u_1)(z_1-\hat z_{2\epsilon})]^\alpha\,,\qquad A_\epsilon \equiv|A_\epsilon^\alpha|\,,\nonumber\\
v_\epsilon^\alpha&=&(P(u_1)u_{2\epsilon})^\alpha\,.
\eea
Here, $P(u)^\alpha{}_\beta=\delta^\alpha_\beta +u^\alpha u_\beta$ is the projector orthogonal to the unit timelike vector $u$, $u \cdot u=-1$.
Since $z_1^\alpha-\hat z_{2\epsilon}^\alpha$ is a null vector
\beq
z^\alpha-\hat z_{2\epsilon}^\alpha=(P(u_1)(z_1-\hat z_{2\epsilon}))^\alpha+u_1^\alpha\, [u_1\cdot (z_1-\hat z_{2\epsilon})]\,,
\eeq
with 
\beq
|(P(u_1)(z_1-\hat z_{2\epsilon}))^\alpha|=|u_1\cdot (z_1-\hat z_{2\epsilon})|\,,
\eeq
that is
\beq
\epsilon \rho_\epsilon A_\epsilon=u_1\cdot (z_1-\hat z_{2\epsilon})\,,
\eeq
where one has taken into account that for the retarded point the component of $(z_1-\hat z_{2\epsilon})$ along $u_1$ is positive whereas
for the advanced point the latter is negative.
Finally, note that the decomposition of the null vector $z_1^\mu-\hat z_{2\epsilon}^\mu$ with respect to the $u_1^\mu$ time axis can be written in the following two ways
\bea
z_1^\mu-\hat z_{2\epsilon}^\mu &=&\rho_\epsilon (A^\mu_\epsilon + \epsilon u_1^\mu  A_\epsilon )\,, \nonumber\\
&=&\rho_\epsilon A_\epsilon ( N_\epsilon^\mu +\epsilon  u_1^\mu   )=B_\epsilon (N_\epsilon^\mu + \epsilon u_1^\mu  )\,,
\eea
where we introduced the unit spatial vector
\beq
 N_\epsilon^\mu \equiv \frac{A_\epsilon^\mu}{A_\epsilon}\,,
\eeq
and the modulus
\beq
B_\epsilon \equiv\rho_\epsilon A_\epsilon\,.
\eeq
\section{Notation, definitions and a list of useful relations}
\label{notation}

Let us consider the definition \eqref{D_and_S_def} of $D(\tau)$ and $S(\tau)$.
A number of useful relations can be found, e.g. $S(\tau)S(-\tau)=1$, as well as
\beq
I_n=\int_{-\infty}^\infty \frac{d\tau}{D^n(\tau)}=\frac{b_0^{1-n}}{\sqrt{\gamma^2-1}}\frac{\Gamma (\frac{n-1}{2})}{\Gamma(\frac{n}{2})}\,.
\eeq
Of special interest  are then the values of the functions $D(\tau'_\epsilon(\tau) )$ and $S(\tau'_\epsilon(\tau))$ where   
\beq
\tau'_\epsilon(\tau)=\gamma \tau -\epsilon D(\tau)\,.
\eeq
We find
\bea
D(\tau'_\epsilon(\tau) )&=& \gamma D(\tau)-\epsilon (\gamma^2-1)\tau=A_\epsilon \rho_\epsilon=B_\epsilon \nonumber\\
S(\tau'_\epsilon(\tau)) &=& \frac{\sqrt{\gamma^2-1}}{b_0}[\gamma \tau -\epsilon D(\tau)]+\frac{A_\epsilon \rho_\epsilon}{b_0}\,,
\eea
where all quantities in these expressions ($\tau'_\epsilon $, $A_\epsilon$,  $\rho_\epsilon$, etc.) are meant to be the corresponding zeroth-PM-order values. 
Similarly,  the following (less evident) relations hold
\bea
S(\gamma\tau+D(\tau))S(\gamma\tau-D(\tau))&=& S(\tau)^2\,,\nonumber\\
D(\gamma\tau+D(\tau))D(\gamma\tau-D(\tau))&=&  
\gamma^2D^2(\tau/\gamma)\,,\nonumber\\
D(\gamma\tau+D(\tau))+D(\gamma\tau-D(\tau))&=&2\gamma D(\tau)\,, \nonumber\\
D(\gamma\tau+D(\tau))-D(\gamma\tau-D(\tau))&=&2\tau (\gamma^2-1)
\,,
\eea
also implying
\bea
D(\gamma\tau+D(\tau))&=& \gamma D(\tau)+\tau (\gamma^2-1)\,,\nonumber\\
D(\gamma\tau-D(\tau))&=&\gamma D(\tau)-\tau (\gamma^2-1)\,,
\eea
and
\bea
&& (\gamma\tau+D(\tau))D(\gamma\tau-D(\tau))\nonumber\\
&& +(\gamma\tau-D(\tau))D(\gamma\tau+D(\tau))=2\tau D(\tau)\,.
\eea

\section{Linear momentum at $O(G)$: computational details}
\label{lin_mom_A}

In Eq. \eqref{P_and_J} (and related Eqs. \eqref{P_and_J_kin} and \eqref{P_and_J_field})  we distinguished a kinematical and an interaction part in the total linear momentum.
A direct evaluation at order $O(G)$ of the kinematical part leads to 
\beq
P_{\rm kin}^\mu(\tau_1,\tau_2)=Ap_1^\mu+Bp_2^\mu+C e_x^\mu\,,
\eeq
with
\bea
A&=&1+Gm_2 \frac{\gamma(2\gamma^2-3)}{\gamma^2-1}\left(-\frac{\gamma}{D(\tau_1)}+\frac{1}{D(\tau_2)}\right) \,,\nonumber\\
B&=&1-Gm_1 \frac{\gamma(2\gamma^2-3)}{\gamma^2-1}\left(\frac{\gamma}{D(\tau_2)}-\frac{1}{D(\tau_1)}\right)\,, \nonumber\\
C&=& -\frac{Gm_1m_2}{b_0}(2\gamma^2-1)\left(\frac{\tau_1}{D(\tau_1)}-\frac{\tau_2}{D(\tau_2)}\right)\,.
\eea
The interaction part instead reduces to
\bea
P_{\rm int}^\mu (\tau_1,\tau_2)&=& Gm_2p_1^\mu \frac{\gamma(2\gamma^2-3)}{(\gamma^2-1)}\left(\frac{\gamma}{D(\tau_1)}-\frac{1}{D(\tau_2)}\right)\nonumber\\
&+& Gm_1p_2^\mu \frac{\gamma(2\gamma^2-3)}{(\gamma^2-1)}\left(\frac{\gamma}{D(\tau_2)}-\frac{1}{D(\tau_1)}\right)\nonumber\\
&+& 2Gm_1m_2 (2\gamma^2-1)(b_1^\mu-b_2^\mu)\tilde I_1(\tau_1,\tau_2)\,,\nonumber\\
\eea
where $b_1^\mu-b_2^\mu=b_0e_x^\mu$ and
\beq
\tilde I_1(\tau_1,\tau_2)=\left(\int_{\tau_1}^{\infty}\int_{-\infty}^{\tau_2}
-\int_{-\infty}^{\tau_1}\int_{\tau_2}^{\infty}\right)d\tau d\bar\tau \delta'(w)\,.
\eeq
Evaluating  $\tilde I_1$ is straightforward and gives
\beq
\tilde I_1(\tau_1,\tau_2)=\frac{1}{2b_0^2}\left(\frac{\tau_1}{D(\tau_1)}-\frac{\tau_2}{D(\tau_2)}\right)\,.
\eeq
Summing the two contributions, kinematical and interaction, one finds that all time dependent terms at $O(G^1)$ cancel and the total mechanical momentum of the system 
$P^{\rm sys}(\tau_1,\tau_2)=P^{\rm kin}(\tau_1,\tau_2)+
P^{\rm int}(\tau_1,\tau_2)$ turns out to be
\beq
P^{\rm sys}(\tau_1,\tau_2)=p_1+p_2+O(G^2)\,,
\eeq
namely is conserved and equal to the initial value. Note, in fact, that the interaction part $P_{\rm int}^\mu (\tau_1,\tau_2)$ vanishes both in the incoming state, and in the outgoing state. By contrast, this property does not
hold for the interaction angular momentum, as discussed next.

\section{Angular momentum at $O(G)$: computational details}
\label{ang_mom_A}

In Eqs. \eqref{P_and_J}, \eqref{P_and_J_kin} and \eqref{P_and_J_field}   we distinguished a kinematical and an interaction part for the total  Noetherian angular momentum of the system.
A direct evaluation at order $O(G)$ of the kinematical part
\beq
J^{\mu\nu}_{\rm kin}=\left[z_1 \wedge m_1 \dot z_1 +z_2 \wedge m_2 \dot z_2  \right]^{\mu\nu}\,,
\eeq
yields
\bea
J^{\mu\nu}_{\rm kin}&=&b_1E_1^{\mu\nu}+b_2 E_2^{\mu\nu}\nonumber\\
&+& G [A m_2 E_1+B m_1 E_2+CE_3]^{\mu\nu}+O(G^2)\,,
\eea
where  we define the bivectors
\bea
E_1^{\mu\nu}& \equiv&(e_x\wedge p_1)^{\mu\nu}=(\hat b_{12}\wedge p_1)^{\mu\nu}\,,\nonumber\\
E_2^{\mu\nu}&\equiv&(e_x\wedge p_2)^{\mu\nu}=(\hat b_{12}\wedge p_2)^{\mu\nu}\,,\nonumber\\ 
E_3^{\mu\nu}&\equiv&(p_1\wedge p_2)^{\mu\nu}\,,
\eea
and where the coefficients read
\bea
A&=&-\frac{2\gamma^2+1}{D(\tau_1)}b_1+\left(\frac{\gamma (2\gamma^2-3)}{D(\tau_2)}+\frac{2\gamma^2-1}{D(\tau_1)}  \right)\frac{b_2}{(\gamma^2-1)}\nonumber\\
&+&\frac{2\gamma^2-1}{\gamma^2-1}\,, \nonumber\\
B&=& \left(\frac{\gamma (2\gamma^2-3)}{D(\tau_1)}+\frac{2\gamma^2-1}{D(\tau_2)}  \right)\frac{b_1}{(\gamma^2-1)} -\frac{2\gamma^2+1}{D(\tau_2)}b_2\nonumber\\
&-& \frac{2\gamma^2-1}{\gamma^2-1}\,,\nonumber\\
C&=& \frac{\gamma (2\gamma^2-3)}{\gamma^2-1}\left(\frac{\tau_1}{D(\tau_1)}-\frac{\tau_2}{D(\tau_2)}\right.\nonumber\\
&-&\left.\frac{1}{\sqrt{\gamma^2-1}}\ln \left(\frac{S(\tau_1)}{S(\tau_2)}\right)\right)\,.
\eea

One sees that the kinematical part of the angular momentum tensor is not constant by itself, as it contains 
contributions depending on  $\tau_1$ and $\tau_2$. Furthermore, the latter  $\tau_a$-dependent contributions
vanish neither in the incoming state, nor in the outgoing one. These are the ``scoot" terms mentioned in the text.
We next show that they are cancelled by corresponding (opposite) contributions contained in the
interaction part of the angular momentum.

We have summarized in Eqs. \eqref{P_and_J_field} above, following Ref. \cite{Friedman:2005rx}, the  \lq\lq int" or \lq\lq Fokker" part of the angular momentum
In these relations, after differentiation with respect to $\dot z_1$, we use the proper time parametrization 
(so that  $\dot z_a \to u_a$) and the $G^0$ \lq\lq straight lines" solution 
\beq
z_1(\tau_1)=b_1 e_x +\bar u_1 \tau_1\,,\qquad z_2(\tau_2)=b_2  e_x +\bar u_2 \tau_2\,,
\eeq
obtaining 
\bea
{\mathcal P}_{1\beta}|_{\rm prop\, time}
&=&  m_1 m_2 \delta(w)\left[-4\gamma \bar u _{2\beta} +(2\gamma^2+1) \bar u_{1\beta}\right]\,,\nonumber\\
\Lambda|_{\rm prop\, time}&=&m_1 m_2 \delta(w)(2\gamma^2-1)\,.
\eea
Therefore
\bea
(z_1\wedge {\mathcal P}_1)(\tau_1,\bar\tau)&=& \delta(w) [(2\gamma^2+1)m_2b_1E_1\nonumber\\
&-&4\gamma m_1b_1 E_2 -4\gamma \tau_1E_3]\,,\nonumber\\
(z_2\wedge {\mathcal P}_2)(\tau,\tau_2) &=& \delta(w) [-4\gamma m_2b_2 E_1\nonumber\\
&+&(2\gamma^2+1)m_1b_2E_2 +4\gamma \tau_2E_3]\,,
\eea
and
\bea
(u_1\wedge {\mathcal P}_1)(\tau,\bar\tau)&=&-4\gamma \delta(w)E_3\,,\nonumber\\
(z_1\wedge {\mathcal Q})(\tau,\bar\tau)&=&-2\frac{\partial \Lambda}{\partial w}(z_1\wedge z_2)\nonumber\\
&=&
-2 (2\gamma^2-1) \delta'(w) m_1 m_2(z_1\wedge z_2)\nonumber\\
&=&
 -2(2\gamma^2-1)\delta'(w)[-m_2b_2 \tau  E_1\nonumber\\
&&+m_1b_1 \bar\tau  E_2+\tau \bar\tau E_3]\,,
\eea
where we used
\beq
m_1 m_2 z_1\wedge z_2=-m_2 b_2 \tau_1 E_1+m_1 b_1 \tau_2 E_2+\tau_1\tau_2 E_3\,.
\eeq
Let us introduce the notation
\bea
\left(\int_{\tau_1}^{\infty}\int_{-\infty}^{\tau_2}
-\int_{-\infty}^{\tau_1}\int_{\tau_2}^{\infty}\right)d\tau d\bar\tau\, \delta(w)&=&I_1(\tau_1,\tau_2)\,,\nonumber\\
\left(\int_{\tau_1}^{\infty}\int_{-\infty}^{\tau_2}
-\int_{-\infty}^{\tau_1}\int_{\tau_2}^{\infty}\right)d\tau d\bar\tau\,  \tau\delta'(w)&=&I_2(\tau_1,\tau_2)\,,\nonumber\\
\left(\int_{\tau_1}^{\infty}\int_{-\infty}^{\tau_2}
-\int_{-\infty}^{\tau_1}\int_{\tau_2}^{\infty}\right)d\tau d\bar\tau\, \bar\tau\delta'(w)&=&I_3(\tau_1,\tau_2)\,,\nonumber\\
\left(\int_{\tau_1}^{\infty}\int_{-\infty}^{\tau_2}
-\int_{-\infty}^{\tau_1}\int_{\tau_2}^{\infty}\right)d\tau d\bar\tau\, \tau\bar\tau \delta'(w)&=&I_4(\tau_1,\tau_2)\,.\nonumber\\
\eea
Changing the names of the integration variables $\tau \leftrightarrow \bar\tau$ , and exchanging $\tau_1$ with $\tau_2$ one immediately has
\beq
\label{I3_I2}
I_3(\tau_1,\tau_2)=-I_2(\tau_2,\tau_1)\,,
\eeq
while, using the formula (D12) of Ref. \cite{Friedman:2005rx}, one has
\beq
\label{I1_from_D12}
I_1(\tau_1,\tau_2)=-\frac{1}{\sqrt{\gamma^2-1}}\ln \left(\frac{S(\tau_1)}{S(\tau_2)}\right)\,.
\eeq
Furthermore we can summarize the integrals $I_n=[I_2,I_3,I_4]$ as
\beq
I_n(\tau_1,\tau_2)=\left(\int_{\tau_1}^{\infty}\int_{-\infty}^{\tau_2}
-\int_{-\infty}^{\tau_1}\int_{\tau_2}^{\infty}\right)d\tau d\bar\tau\,f_n(\tau, \bar\tau) \delta'(w) \,, 
\eeq
with $n=[2,3,4]$ and correspondingly
\beq
f_n(\tau, \bar\tau)=[\tau, \bar\tau, \tau \bar\tau]\,.
\eeq 
The final result for the interaction part of the angular momentum reads
\begin{widetext}
\bea
J^{\mu\nu}_{\rm int}(\tau_1,\tau_2)&=&m_2\left[(2\gamma^2+1)\frac{b_1}{D(\tau_1)}-4\gamma \frac{b_2}{D(\tau_2)}+2b_2(2\gamma^2-1)I_2(\tau_1,\tau_2)  \right]E_1^{\mu\nu}\nonumber\\
&+&m_1\left[(2\gamma^2+1)\frac{b_2}{D(\tau_2)}-4\gamma \frac{b_1}{D(\tau_1)}-2b_1(2\gamma^2-1)I_3(\tau_1,\tau_2)  \right]E_2^{\mu\nu}\nonumber\\
&+& \left[-4\gamma \frac{\tau_1}{D(\tau_1)}+4\gamma \frac{\tau_2}{D(\tau_2)}-4\gamma I_1(\tau_1,\tau_2)-2(2\gamma^2-1)I_4(\tau_1,\tau_2)  \right]E_3^{\mu\nu}\,.
\eea
\end{widetext}
A direct evaluation of these integrals gives
\bea
2(\gamma^2-1)I_2(\tau_1,\tau_2)&=&-\frac{1}{D(\tau_1)}+\frac{\gamma}{D(\tau_2)}\,,\nonumber\\
2(\gamma^2-1)I_3(\tau_1,\tau_2)&=&\frac{1}{D(\tau_2)}-\frac{\gamma}{D(\tau_1)}\,,\nonumber\\
2\frac{(\gamma^2-1)}{\gamma}I_4 &=&-\frac{\tau_1}{D(\tau_1)}+\frac{\tau_2}{D(\tau_2)}\nonumber\\
&+&\frac{1}{\sqrt{\gamma^2-1}}\ln \left(\frac{S(\tau_1)}{S(\tau_2)}\right)\nonumber\\
&=&-\frac{\tau_1}{D(\tau_1)}+\frac{\tau_2}{D(\tau_2)}-I_1(\tau_1,\tau_2) \,,\nonumber\\
\eea
that is
\beq
I_3(\tau_1,\tau_2)=-I_2(\tau_2,\tau_1)\,,
\eeq
and one must recall the result \eqref{I1_from_D12} for $I_1(\tau_1,\tau_2)$, namely
\beq
I_1(\tau_1,\tau_2)=-\frac{1}{\sqrt{\gamma^2-1}}\ln \left(\frac{S(\tau_1)}{S(\tau_2)}\right)\,.
\eeq
When considering the (incoming or outgoing) asymptotic values of $J^{\mu\nu}_{\rm int}$, i.e., taking the limits $\tau_1\sim \tau_2$ going to $-\infty$ or 
$\tau_1\sim \tau_2$ going to $\infty$ we find that $J^{\mu\nu}_{\rm int}$ {\it does not tend to zero} in the asymptotic region. Defining
\beq
\label{small_j_mu_nu}
j^{\mu\nu}=G^2 m_1 m_2 \frac{\gamma(2\gamma^2-3)}{(\gamma^2-1)^{3/2}}\, (u_1\wedge u_2)^{\mu\nu}\,,
\eeq
we find
\bea
\label{j_int_plus}
(J^{\mu\nu}_{\rm int})^{+\infty}&=&j^{\mu\nu}\ln \frac{|\tau_1^+|}{|\tau_2^+|}=j^{\mu\nu}\ln \frac{\sqrt{1-({\mathbf v}_1^+)^2}}{\sqrt{1-({\mathbf v}_2^+)^2}}
\nonumber\\
&=&j^{\mu\nu}\ln \frac{E_2^+}{E_1^+}\,,
\eea
while 
\bea
\label{j_int_minus}
(J^{\mu\nu}_{\rm int})^{-\infty}&=&-j^{\mu\nu}\ln \frac{|\tau_1^-|}{|\tau_2^-|}=-j^{\mu\nu}\ln \frac{\sqrt{1-({\mathbf v}_1^-)^2}}{\sqrt{1-({\mathbf v}_2^-)^2}}\nonumber\\
&=&-j^{\mu\nu}\ln \frac{E_2^-}{E_1^-}\,.
\eea
Here we consider the asymptotic limit in some Lorentz frame, i.e., $\tau_a^\pm \approx \pm \sqrt{1-({\mathbf v}_a^\pm)^2}|t|$, and $E_a^\pm=\frac{m_a}{\sqrt{1-({\mathbf v}_a^\pm)^2}}$ where $({\mathbf v}_a^\pm)$ denotes the asymptotic velocities  of particle $a$. 

The asymptotic contributions $\propto j^{\mu\nu}$ in $J^{\mu\nu}_{\rm int}$ precisely cancell the
corresponding scoot contributions in $J^{\mu\nu}_{\rm kin}$.
Summing the $O(G^1)$ kinematical and interaction parts we find the result given in Eq. \eqref{J_sys_final}.

\end{document}